\documentclass[aps,prl,reprint,superscriptaddress]{revtex4-2}

\usepackage{graphicx}
\usepackage{amsfonts}
\usepackage{color}
\usepackage{xcolor}

\usepackage[normalem]{ulem}

\begin{document}

\title{Resolving Structural Avalanches in Amorphous Carbon with Arclength Continuation}

\author{Fraser Birks}
\email{fraser.birks@warwick.ac.uk}
\affiliation{Warwick Centre for Predictive Modelling, School of Engineering, University of Warwick, Coventry CV4 7AL, UK}

\author{Ibrahim Ghanem}
\author{Lars Pastewka}
\affiliation{Department of Microsystems Engineering, University of Freiburg, Georges-K\"ohler-Allee 103, 79110 Freiburg, Germany}

\author{James Kermode}
\affiliation{Warwick Centre for Predictive Modelling, School of Engineering, University of Warwick, Coventry CV4 7AL, UK}

\author{Maciej Buze}
\affiliation{MARS: Mathematics for AI in Real-world Systems, School of Mathematical Sciences, Lancaster University, Lancaster, LA1 4YF, UK}

\date{\today}
\begin{abstract}
Plastic deformation in amorphous solids is carried by localized shear transformations that self-organize into avalanches. In amorphous carbon modeled with a machine-learned interatomic potential, we find that the energetics and organization of these avalanches can be resolved by systematically following the underlying energy landscape. With a pseudo-arclength numerical continuation framework, we decompose avalanches into constituent shear transformations and determine their strain-dependent energetics. Our analysis shows that, prior to onset, avalanches have a latent structure that consists of well-separated local minima. We further demonstrate that arclength continuation yields an event driven framework for following avalanche dynamics, eliminating time-step effects on statistical avalanche properties such as distributions of stress drops.
\end{abstract}

\maketitle

During plastic deformation, materials undergo irreversible changes in their structure.
In crystalline solids, the mechanism underlying this irreversibility is well established: it is the glide of dislocations, line defects whose motion results in the net sliding of planes of atoms~\cite{hirth_theory_1983}.
In contrast, glassy materials, which lack long-range structural order, deform plastically through the localized yielding of small regions within the amorphous structure~\cite{argon_plastic_1979,spaepen_microscopic_1977}.
These ``soft'' regions, referred to as shear transformation zones (STZs), typically consist of a few tens of atoms that undergo highly non-affine rearrangements~\cite{falk_dynamics_1998}.

Extensive research has been devoted to understanding the fundamental nature of STZs and determining whether there are structural indicators that can reliably predict their emergence~\cite{richard_predicting_2020}.
A major challenge in this endeavor lies in the fact that STZs are not isolated entities, they interact through long-range elastic fields~\cite{maloney_subextensive_2004}.
As a result, the activation of a single STZ can trigger the yielding of nearby regions, potentially leading to a cascade of correlated events, or structural avalanches~\cite{sethna_crackling_2001}.
Avalanches are most pronounced at low temperature and are often studied in athermal and quasi-static (AQS) simulations~\cite{Maloney2006PRE}.
The phenomenology of systems at finite-temperature and finite rate is similar, with temperature and rate cutting off the observable avalanche sizes at small and large scales, respectively~\cite{Caroli_rate,Edan_temp,Liu2016-tc,phase_diagram}.

\begin{figure}
    \centering
    \includegraphics[width=\linewidth]{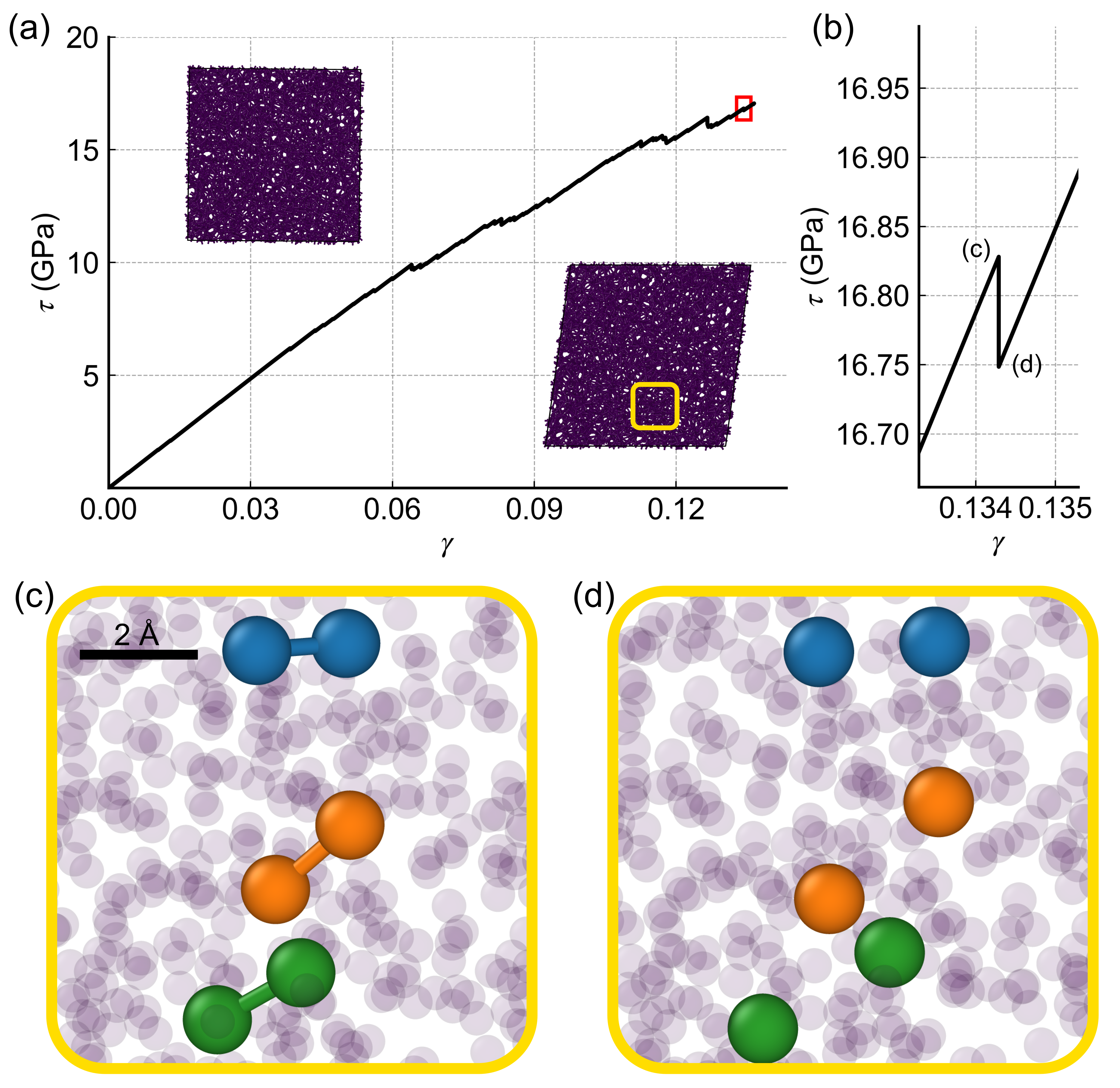}
    \caption{Structure and plastic events in amorphous carbon. (a) The shear-stress shear-strain curve of a 4096-atom structure of amorphous carbon with a density of 2.7~$\text{gcm}^{-3}$. Note that $\gamma$ is the simple shear strain. (b) An enlargement of the stress-strain curve corresponding to the position of the red box in (a). The stress drop encircled corresponds to a three-bond structural avalanche plastic event. (c-d) The 3-bonds involved in the structural avalanche highlighted in (b), both before (c) and after (d) the event was triggered. The three bonds involved are colored blue, orange and green.}
    \label{fig1}
\end{figure}

AQS uses sequences of finite-strain increments together with searches for the next local minimum to trace out the most likely low-temperature path of the driven system within its potential energy landscape.
Fig.~\ref{fig1} demonstrates the AQS method for computing a stress--strain curve for the simple shear of a 4096 atom sample of amorphous carbon (a-C, density 2.7~$\text{g\,cm}^{-3}$), generated via a rapid liquid quench of carbon from 10,000~K at a rate of 1000~K/ps. For interatomic interactions, we used the realistic machine-learned atomic cluster expansion (ACE) potential from Qamar et al. \cite{qamar2023atomic,note1}.\
The stress-strain curve is shown in Fig.~\ref{fig1}a. Discontinuous stress drops that can be seen on this curve correspond to the activation of plastic events. Due to the covalent nature of bonding in a-C, plastic events necessarily involve the breaking and/or formation of bonds \cite{Kunze_2014}. While some stress drops are associated with a single bond rearrangement, the majority are structural avalanches involving multiple bonds. One such example is the three-bond structural avalanche highlighted in Figs.~\ref{fig1}c and d. 

Two properties of AQS limit its usefulness in the study of avalanche events: (i) The finite step size makes it difficult to determine whether events are truly correlated, or just occurring at strain values too close to resolve, and (ii) AQS simulations give no information about the inherent energetics of an avalanche, other than the trivial fact that the first barrier is zero at the instant of onset.

One way of overcoming shortcoming (ii) is to identify two mechanically stable configurations at the same applied strain before and after instability onset, and perform a nudged elastic band (NEB)~\cite{Henkelman2000CI-NEB, Henkelman2000NEB} calculation to determine the minimum energy path.
This approach was employed in Ref.~\cite{Rodney2011MSMSE} to compute the energy barriers associated with instabilities for both single shear transformations and avalanches.

We have found that while using NEB directly can work for small avalanches with fewer than three  bonds involved, it is completely unsuitable for larger avalanches. There are three reasons for this: (i) Many images (often $>150$) are required to adequately resolve energy landscapes containing multiple intermediate minima and barriers of different sizes. (ii) Convergence frequently fails when near the avalanche onset strain. (iii) The energy pathway determined (the ordering of events) is highly sensitive to the initial guess. For further detail, see section A of the Supplementary Material \cite{SupplementalA}.

In this Letter, we introduce a tailored numerical arclength continuation (AC) method which overcomes the limitations of both AQS and NEB simulations.
With this method, we remove step-size dependence and show that avalanches would happen in an AQS simulation even with an infinitesimally small strain step. We reveal that at strains below onset, avalanches in amorphous carbon modeled with a realistic potential decompose into latent structures comprising chains of well separated local minima connected by index-1 saddle points.
Going further, with minimal effort, we obtain full energetic information between the intermediate basins and saddle points of the latent structure at a range of strains. Once these are identified, we validate our results by running NEB calculations between adjacent intermediate states. 

AC is a method for tracking continuous solution branches of non-linear equations as a control parameter varies \cite{allgower2003introduction}.
In the context of amorphous plasticity, we are interested in finding sets of atomic positions $\mathbf{R}\, \in \mathbb{R}^{3N}$(where $N$ is the number of atoms) that correspond to energy minima or saddle points of a system, where all atomic forces vanish $\mathbf{F}(\mathbf{R},\gamma)=\mathbf{0}$ simultaneously, as the applied strain $\gamma$ changes. 
The key idea of AC is to parameterize the solution branch $\{(\mathbf{R}(s), \gamma(s))\}_s$ using an arclength parameter $s$.
In the AQS approach, the solution path $\{(\mathbf{R}(\gamma),\gamma)\}_{\gamma}$ is always parameterized by the applied strain, leading to singularities at bifurcation points, which are the critical values of the applied strain where STZs activate. 
In contrast, the AC approach allows a seamless traversal of the bifurcation points, meaning it is able to transition smoothly from minima to saddle (provided the interatomic potential is twice continuously differentiable in the vicinity of the bifurcation point, as is expected for any reasonable potential).
The framework also admits a natural adaptive step size in $s$, which avoids the fixed step-size compromise inherent to AQS simulations and ensures that bifurcation events along the solution branch are not missed.

One limitation of typical AC which proves problematic in this application is the requirement for second derivative (Hessian) information. 
For atomistic systems, routes toward computing the Hessian are either numerical or analytical, which are both prohibitively computationally expensive for realistic interatomic potentials such as the ACE potential employed here.  
We therefore apply a Hessian-free pseudo-arclength numerical continuation algorithm similar to that previously applied to atomistic systems by Buze \& Kermode \cite{buze2021numerical}, with full algorithmic details presented in section B of the Supplementary Material \cite{SupplementalB}.
AC-based approaches in atomistic modeling have also previously been explored in Refs.~\cite{li2013bifurcation,buze2020analysis} to study brittle fracture and in Ref.~\cite{pattamatta2014mapping} to study response of an FCC nanoslab of nickel to loading. Ref.~\cite{maliyov2025exploring} also presents a related idea, the use of implicit differentiation to explore atomistic energy landscapes via a Hessian-free approximation.

\begin{figure}
    \centering
    \includegraphics[width=1\linewidth]{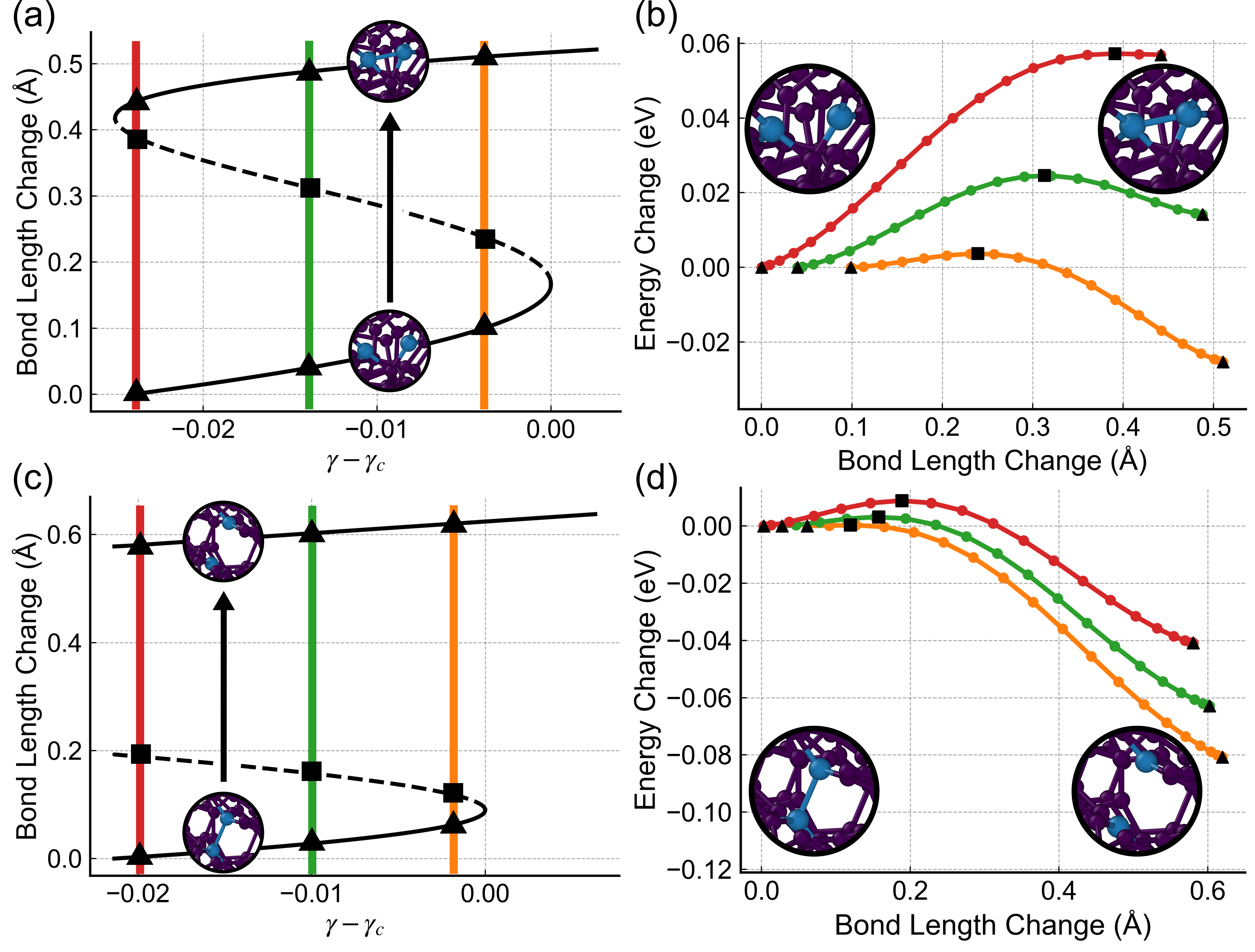}
    \caption{The strain-dependent energy landscape of single-bond plastic events is explored using AC. (a) For a reversible event, the solution branch follows the minimum associated with the open bond (lower segment, solid black) until the critical strain $\gamma_c$, where the barrier vanishes. AC then passes through the bifurcation to the central segment, corresponding to the index-1 saddle traced as strain decreases (dashed black), before reaching the upper segment associated with the formed bond (solid black). (b) Nudged elastic band (NEB) energy profiles sampled across this landscape, with color-coded strains indicated in (a), show excellent agreement between NEB and continuation saddle energies (triangles and squares denote minima and saddles). (c)–(d) The same analysis for an irreversible event shows no continuous  branch from the central to the upper segment, consistent with the strongly asymmetric barrier revealed by the NEB profiles.}
    \label{fig2}
\end{figure}

We first validated this method through application to single-bond plastic events.
In this system, single bond plastic events could be broadly divided into two types; those which reversed upon strain removal (reversible events), and those which did not (irreversible events).
Fig.~\ref{fig2}a shows the branch traversed by the continuation method for a reversible event in which a single bond forms.
The solution branch has three segments. The first (lower) segment, represents an open bond, with the rightmost point corresponding to the strain at which the bond would naturally form in a quasi-static simulation. 
After the first bifurcation point, the central segment (dashed line), where strain values retreat, represents the set of saddle points which lie between the two stable segments.
Finally, after the second bifurcation point, the system re-enters a minimum, with increasing strain corresponding to the further closing of the two (now bonded) atoms.
Three nudged-elastic-band (NEB) calculations across this bifurcation diagram are shown in Fig.~\ref{fig2}b. The strain value of each NEB corresponds to the vertical line of matching color across Fig.~\ref{fig2}a.
The energies from the continuation branch minima are marked as triangles, while the saddle energies predicted by the continuation are marked as squares. 

Fig.~\ref{fig2}c and d demonstrate how this technique can be applied to an irreversible event. In this case, it can be seen that there is not a single continuous  branch connecting stable segments.
The initial procedure is the same, but rather than traversing the full branch with continuation directly, a relaxation is carried out from the saddle into the adjacent minima, effectively `jumping' the gap.
By inspecting the energy landscape in Fig.~\ref{fig2}d it is clear why such a gap exists --- the energy barrier associated with the event is greatly asymmetrical, and remains similarly asymmetrical for a range of strains.
In both the reversible and irreversible cases, the agreement between the NEB and continuation saddle energies is excellent.
\begin{figure}
    \centering
    \includegraphics[width=1\linewidth]{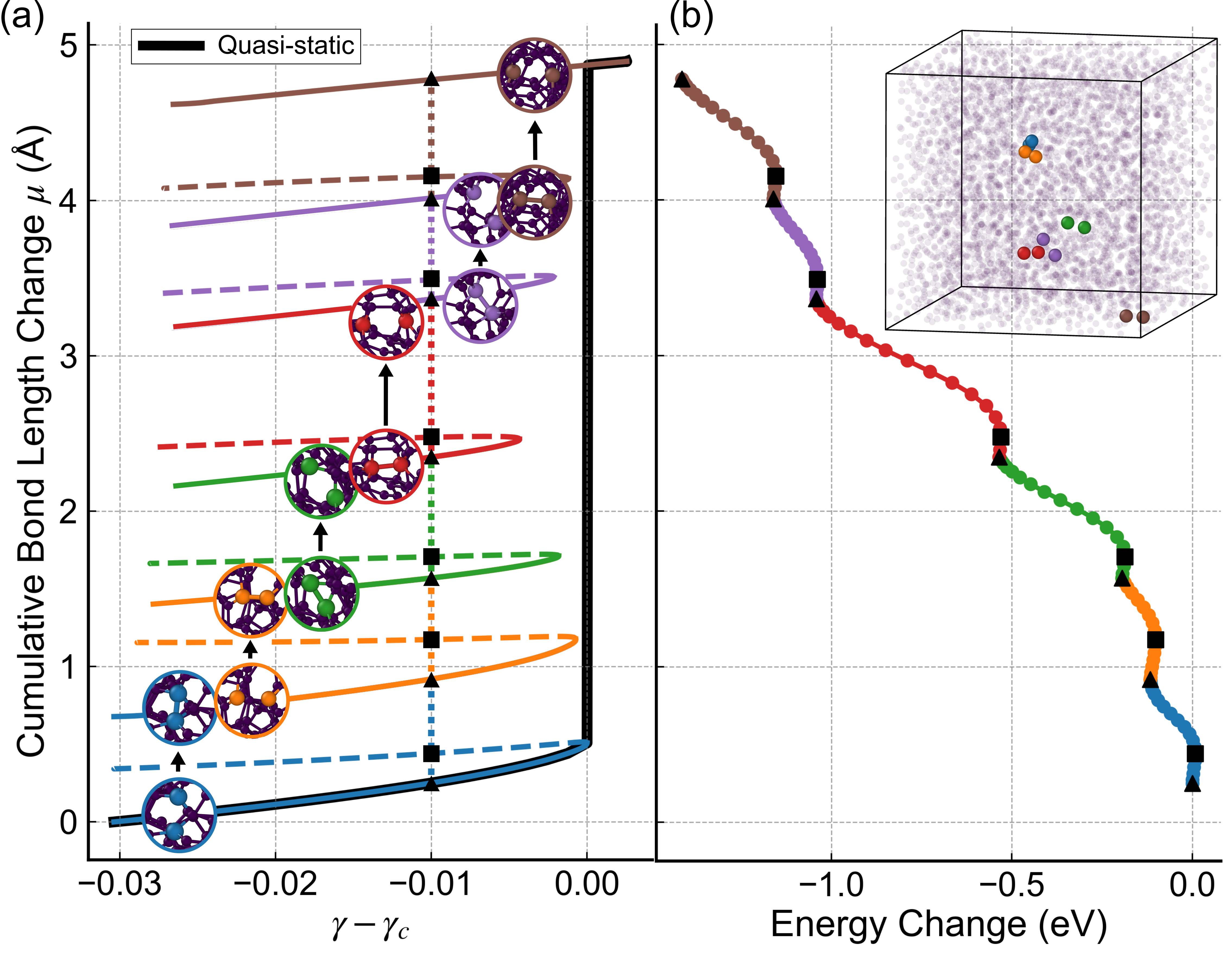}
    \caption{The strain-dependent energy landscape of a six-bond avalanche. (a) A set of continuation branches reveals the latent structure of the avalanche. The AQS trajectory (thick black line) triggers the full avalanche at the critical strain $\gamma_c$. In contrast, AC can traverse the bifurcation at $\gamma_c$ onto the index-1 saddle associated with the first bond-breaking event (dashed blue). Relaxation from this saddle at lower strain leads to an intermediate basin in which a single bond has broken (orange), and repeating this procedure (green, red, purple, and brown) resolves the avalanche into successive events. All intermediate basins disappear below $\gamma_c$, implying that even an infinitesimal-step AQS protocol would still trigger the full avalanche. (b) A nudged elastic band (NEB) energy landscape at $\gamma_{xy}-\gamma_c=-0.01$, assembled from sequential NEBs between adjacent intermediate basins, with continuation minima and saddles marked by triangles and squares. The NEB and continuation saddle energies agree closely. Insets in (a) and (b) indicate the ordering of events and the full avalanche.}
    \label{fig3}
\end{figure}

We now turn to the complex case of traversing a multi-event avalanche with AC.
For visualization, we define a reaction coordinate, the cumulative active bond-length change $\mu$, defined as $\sqrt{\sum_{ij}(r_{ij}-r^{0}_{ij})^{2}},$ where $r$ and $r^0$ refer to bond lengths in the deformed and a reference structure respectively and the sum runs over the `active' bonds between pairs of atoms $i,j$ (all bonds involved in the avalanche event).
Avalanche exploration proceeds through a controlled sequence of continuation and relaxation steps.
As in the irreversible single-bond case, we first run continuation through the initial bifurcation point to the first saddle. Biased relaxations are performed from the saddle into the adjacent basins.
Near the avalanche onset strain, the system undergoes the full cascade, but at lower strains the system lands in intermediate basins corresponding to triggering only a subset of events. For the six-bond avalanche explored in Fig.~\ref{fig3}a, relaxing from the first saddle (blue dashed line) to the adjacent minima (lower orange solid line) corresponds to the triggering of only a single event.
By iterating this procedure --- starting from each newly discovered basin and using continuation to traverse the next saddle --- we can reconstruct the avalanche as a chain of discrete transitions. In Fig.~\ref{fig3}a, the six-event avalanche has been dissected into six unique single-bond events, colored blue, orange, green, red, purple and brown. A full energy profile across these events formed from individual NEBs between each adjacent intermediate basin at a strain 0.01 below the onset strain $\gamma_c$ is shown in Fig.~\ref{fig3}b. For all transitions, the continuation barrier energies agree perfectly with those predicted by each NEB calculation.

In general, we observe that avalanches cannot always be split perfectly into sequences of single bond events; it is common for single-bond events to be so correlated that they do not disentangle even at lower strains. The range of avalanche sizes examined, together with a discussion of their degree of separability and algorithmic limitations is provided in section C of the Supplementary Material \cite{SupplementalC}

Our observations highlight three interesting features of avalanches in amorphous carbon: (i) avalanches are formed from distinct sequences of low-dimensional events, (ii) these events are highly separable at strains below avalanche onset and (iii) the sequence displays a robust internal latent structure in which the constituent events are connected by a unique chain of local minima and index-1 saddle points. 
Crucially, this structure only appears below the avalanche onset strain, meaning even an infinitesimal quasi-static strain step taken at the critical strain of the first event would trigger the entire avalanche.

\begin{figure}
    \centering
    \includegraphics[width=1\linewidth]{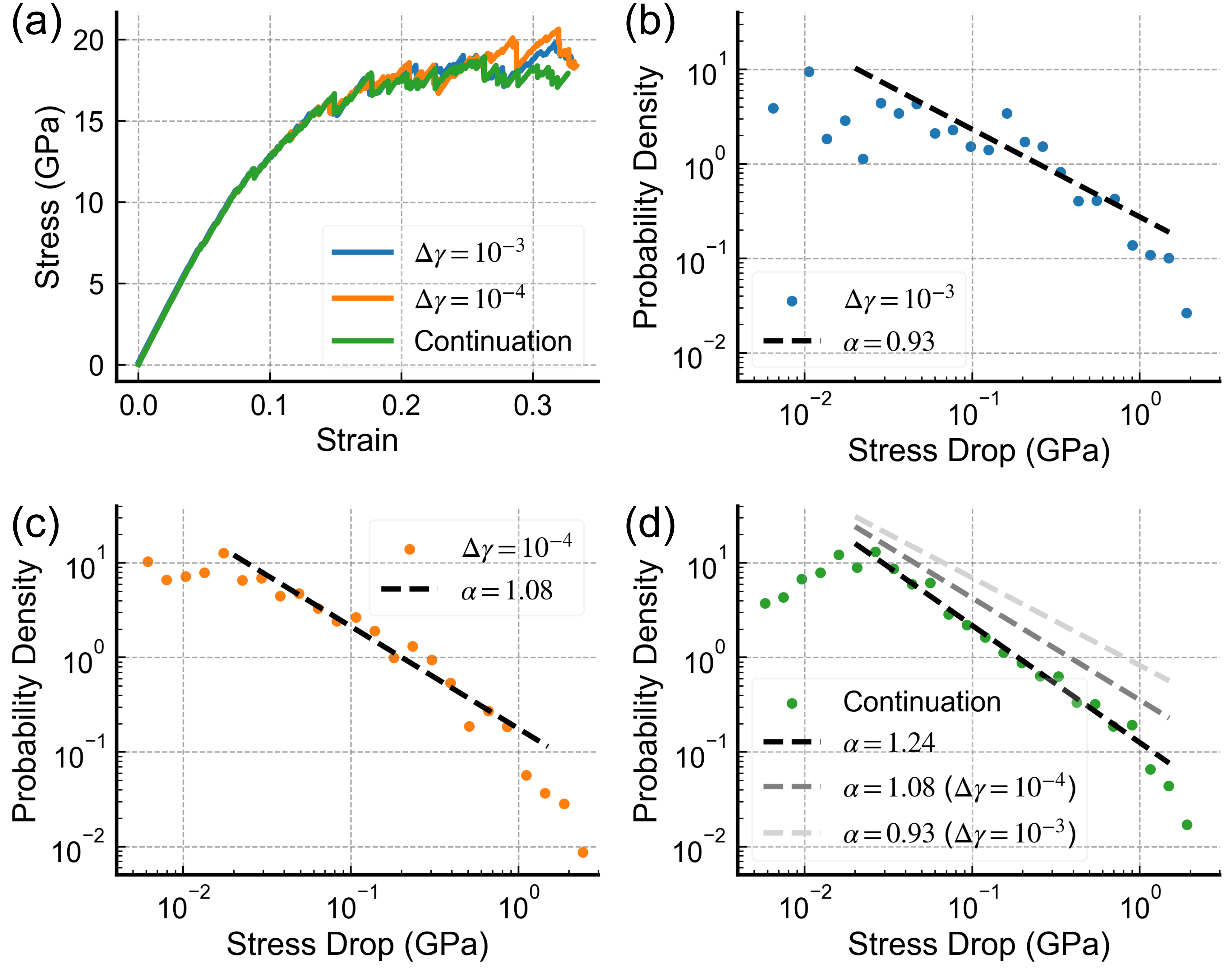}
    \caption{A comparison between AQS and continuation simulations is shown. (a) Overlaid stress–strain curves from AQS simulations with step sizes $\Delta\gamma=10^{-3}$ (blue) and $10^{-4}$ (orange), together with continuation results (green). The curves diverge upon entering the plastic flow regime ($\gamma \gtrsim 0.17$). (b–d) Log–log plots of the binned probability density of stress drops for $\gamma>0.17$ over 5 repeats. AQS results are shown for $\Delta\gamma=10^{-3}$ in (b) and $\Delta\gamma=10^{-4}$ in (c); in both cases, finite step size leads to an underestimation of the exponent $\alpha$ of the fitted power-law $p(\Delta \sigma) \sim (\Delta \sigma)^{-\alpha}$ (dashed black). In contrast, the continuation algorithm in (d) eliminates step-size dependence, yielding a steeper and more clearly defined power-law scaling.}
    \label{fig4}
\end{figure}

Going further, Fig.~\ref{fig4} demonstrates that continuation can not only resolve individual avalanche structures, but also yields a framework for event-driven simulation of avalanche dynamics. Panel (a) compares stress–strain curves obtained from AQS simulations with two different strain increments to the continuation result. While all curves coincide in the elastic regime, they begin to diverge upon entering steady plastic flow ($\gamma \gtrsim 0.17$), reflecting the accumulation of step-size–dependent errors in AQS. Panels (b) and (c) show the corresponding stress-drop distributions measured from a number of AQS trajectories in the plastic flow regime at $\Delta\gamma=10^{-3}$ and $10^{-4}$, respectively. In both cases, the finite strain increment causes multiple nearby instabilities to be triggered within a single step, merging distinct plastic events and suppressing small stress drops. This manifests as a systematic flattening of the distribution and an apparent reduction of the exponent $\alpha$ of the fitted power-law $p(\Delta \sigma) \sim (\Delta \sigma)^{-\alpha}$, which persists even as the step size is reduced. By contrast, panel (d) shows the distribution obtained from AC, where individual bifurcations are resolved directly. The resulting stress-drop statistics are independent of strain step size and exhibit a steeper, more clearly defined power-law scaling, indicating that continuation provides direct access to the intrinsic avalanche statistics underlying plastic flow. The step-size independent nature of the AC method has an additional effect; it allows stress--strain curves to be reproducible with respect to small perturbations in the initial strain. This is demonstrated in Supplementary Material D~\cite{SupplementalD}, together with an analysis of how the computational cost of AC scales with system size and how it compares with the cost of AQS simulations.

The ability to traverse complex regions of the energy landscape for a system modeled with a high-fidelity machine-learned potential suggests that this framework is well suited to interrogate the microscopic physics of plasticity in realistic amorphous systems. In particular, as the energies of minima and saddle segments are directly attainable from continuation, this method opens the way to directly measuring how energy barriers in the system are altered by nearby plastic events; information that could be used to build accurate coarse-grained elasto-plastic models or derive statistical relationships.
Applying AC methods to avalanches in other glassy systems is a natural avenue for future work. Earlier studies—both in simple Lennard–Jones glasses \cite{Salerno_2012,Moriel_2024,Oyama_2024,Dattani_2022} and in metallic glasses using embedded-atom potentials \cite{Duan_2023} could have benefited from this approach.

Beyond these directions, the approach also has clear potential for studying more complex phenomena in amorphous carbon, including fracture processes and the emergence of shear bands, where detailed knowledge of the underlying energy landscape could prove especially informative. 
It would also be valuable to relate the ordering of events identified within an avalanche’s latent structure to that obtained from established avalanche discretization methods, such as overdamped dynamics \cite{Salman_2011,Baggio_2023}. Our preliminary analysis (see Supplementary Material, Section E \cite{SupplementalE}) indicates that although the two orderings are correlated, the dynamical causal chain does not always coincide exactly with the event ordering inferred from the latent structure.

In summary, we have introduced an arclength continuation-based method and applied it to investigate plastic events in amorphous carbon modeled with a machine-learned ACE potential. Unlike traditional approaches combining AQS and NEB simulations, AC is guaranteed to find bifurcation points and can directly traverse saddle points, allowing the extraction of energy barriers for each plastic event.
For this realistic system, we showed that at strains below onset, avalanches can be decomposed into latent structures comprising chains of well-separated local minima connected by index-1 saddle points, the energetics of which were validated separately using NEB calculations between intermediate basins.
Going further, we show that due to the elimination of step-size dependence, the adaptive AC method is particularly well suited to investigating the stress-strain response and statistics of disordered systems with plastic events that are separated by very small strain intervals.
In future, we believe that AC will be a crucial tool in the investigation of avalanches in atomistic systems, and that elucidating avalanche latent structures will allow for a clearer mapping between individual rearrangements, their mutual interactions, and the collective mechanical behavior that emerges at larger scales.

\emph{Acknowledgements --} We are grateful to Matous Mrovec for sharing an early version of the a-C potential published in Ref.~\cite{qamar2023atomic}, and to Albert P. Bartók for helpful discussions. FB is supported by a studentship from the UK Engineering and Physical Sciences Research Council–funded Centre for Doctoral Training in Modelling of Heterogeneous Systems (Grant No. EP/S022848/1). This work was partially funded by the German Research Foundation (DFG) through Research Unit 5099 (Project 431945604). We acknowledge the University of Warwick Scientific Computing Research Technology Platform for computational support. Additional computations were carried out on bwForCluster NEMO2 (State of Baden-Württemberg and DFG Project 455622343), with further resources provided by Sovereign AI and Isambard-AI.
MB is supported by Research England under the Expanding Excellence in England (E3) funding stream, which was awarded to MARS: Mathematics for AI in Real-world Systems in the School of Mathematical Sciences at Lancaster University. Part of this research was performed while MB was a visiting fellow at the Institute for Pure and Applied Mathematics (IPAM), as part of the long program New Mathematics for the Exascale: Applications to Materials Science. IPAM is supported by the U.S. National Science Foundation (Grant No. DMS-1925919).

\emph{Data Availability Statement —} The scripts and data generated for this study are available on Zenodo \cite{Birks2026ArclengthData}. The arclength continuation algorithms used in this work are implemented in LACT, a Python--based LAMMPS \cite{LAMMPS} wrapper, which can be found on GitHub \cite{Buze_LACT_LAMMPS_Continuation}.

\bibliography{refs.bib}

@article{maliyov2025exploring,
    author = "Maliyov, I. and Grigorev, P. and Swinburne, T. D",
    title = "Exploring parameter dependence of atomic minima with implicit differentiation",
    journal = "npj Comput. Mater.",
    volume = "11",
    number = "1",
    pages = "22",
    year = "2025",
    publisher = "Nature Publishing Group UK London",
    doi = "10.1038/s41524-024-01506-0"
}

@Article{LAMMPS,
    author = "Thompson, A. P. and Aktulga, H. M. and Berger, R. and Bolintineanu, D. S. and Brown, W. M. and Crozier, P. S. and in 't Veld, P. J. and Kohlmeyer, A. and Moore, S. G. and Nguyen, T. D. and Shan, R. and Stevens, M. J. and Tranchida, J. and Trott, C. and Plimpton, S. J.",
    title = "{LAMMPS} - a flexible simulation tool for particle-based materials modeling at the atomic, meso, and continuum scales",
    journal = "Comput. Phys. Commun.",
    volume = "271",
    pages = "108171",
    year = "2022",
    doi = "10.1016/j.cpc.2021.108171"
}

@article{buze2020analysis,
    author = "Buze, M. and Hudson, T. and Ortner, C.",
    title = "Analysis of cell size effects in atomistic crack propagation",
    journal = "ESAIM: Mathematical Modelling and Numerical Analysis",
    volume = "54",
    number = "6",
    pages = "1821--1847",
    year = "2020",
    publisher = "EDP Sciences",
    doi = "10.1051/m2an/2020005"
}

@article{pattamatta2014mapping,
    author = "Pattamatta, S. and Elliott, R. S and Tadmor, E. B",
    title = "Mapping the stochastic response of nanostructures",
    journal = "Proc. Natl. Acad. Sci. U.S.A.",
    volume = "111",
    number = "17",
    pages = "E1678--E1686",
    year = "2014",
    publisher = "National Academy of Sciences",
    doi = "10.1073/pnas.1402029111"
}

@article{li2013bifurcation,
    author = "Li, X.",
    title = "A bifurcation study of crack initiation and kinking",
    journal = "Eur. Phys. J. B",
    volume = "86",
    number = "6",
    pages = "258",
    year = "2013",
    publisher = "Springer",
    doi = "10.1140/epjb/e2013-40145-9"
}

@article{buze2021numerical,
    author = "Buze, M. and Kermode, J. R",
    title = "Numerical-continuation-enhanced flexible boundary condition scheme applied to mode-I and mode-III fracture",
    journal = "Phys. Rev. E",
    volume = "103",
    number = "3",
    pages = "033002",
    year = "2021",
    publisher = "APS",
    doi = "10.1103/physreve.103.033002"
}

@book{allgower2003introduction,
    author = "Allgower, E. L and Georg, K.",
    title = "Introduction to numerical continuation methods",
    year = "2003",
    publisher = "SIAM",
    doi = "10.1137/1.9780898719154"
}

@article{qamar2023atomic,
    author = "Qamar, M. and Mrovec, M. and Lysogorskiy, Y. and Bochkarev, A. and Drautz, R.",
    title = "Atomic cluster expansion for quantum-accurate large-scale simulations of carbon",
    journal = "J. Chem. Theory Comput.",
    volume = "19",
    number = "15",
    pages = "5151--5167",
    year = "2023",
    publisher = "ACS Publications",
    doi = "10.1021/acs.jctc.2c01149"
}

@article{hirth_theory_1983,
    author = "Hirth, J. P. and Lothe, J. and Mura, T.",
    title = "Theory of {Dislocations} (2nd ed.)",
    volume = "50",
    issn = "0021-8936, 1528-9036",
    doi = "10.1115/1.3167075",
    number = "2",
    urldate = "2025-08-07",
    journal = "J. Appl. Mech.",
    year = "1983",
    pages = "476--477"
}

@article{argon_plastic_1979,
    author = "Argon, A. S.",
    title = "Plastic deformation in metallic glasses",
    volume = "27",
    copyright = "https://www.elsevier.com/tdm/userlicense/1.0/",
    issn = "00016160",
    doi = "10.1016/0001-6160(79)90055-5",
    number = "1",
    urldate = "2025-08-08",
    journal = "Acta Metal.",
    year = "1979",
    pages = "47--58"
}

@article{falk_dynamics_1998,
    author = "Falk, M. L. and Langer, J. S.",
    title = "Dynamics of viscoplastic deformation in amorphous solids",
    volume = "57",
    copyright = "http://link.aps.org/licenses/aps-default-license",
    issn = "1063-651X, 1095-3787",
    doi = "10.1103/PhysRevE.57.7192",
    number = "6",
    urldate = "2025-08-08",
    journal = "Phys. Rev. E",
    year = "1998",
    pages = "7192--7205"
}

@article{richard_predicting_2020,
    author = "Richard, D. and Ozawa, M. and Patinet, S. and Stanifer, E. and Shang, B. and Ridout, S. A. and Xu, B. and Zhang, G. and Morse, P. K. and Barrat, J.-L. and Berthier, L. and Falk, M. L. and Guan, P. and Liu, A. J. and Martens, K. and Sastry, S. and Vandembroucq, D. and Lerner, E. and Manning, M. L.",
    title = "Predicting plasticity in disordered solids from structural indicators",
    volume = "4",
    issn = "2475-9953",
    doi = "10.1103/PhysRevMaterials.4.113609",
    number = "11",
    urldate = "2025-08-08",
    journal = "Phys. Rev. Mater.",
    year = "2020",
    pages = "113609"
}

@article{maloney_subextensive_2004,
    author = "Maloney, C. and Lemaître, A.",
    title = "Subextensive {Scaling} in the {Athermal}, {Quasistatic} {Limit} of {Amorphous} {Matter} in {Plastic} {Shear} {Flow}",
    volume = "93",
    copyright = "http://link.aps.org/licenses/aps-default-license",
    issn = "0031-9007, 1079-7114",
    doi = "10.1103/PhysRevLett.93.016001",
    number = "1",
    urldate = "2025-08-08",
    journal = "Phys. Rev. Lett.",
    year = "2004",
    pages = "016001"
}

@article{sethna_crackling_2001,
    author = "Sethna, J. P. and Dahmen, K. A. and Myers, C. R.",
    title = "Crackling noise",
    volume = "410",
    copyright = "http://www.springer.com/tdm",
    issn = "0028-0836, 1476-4687",
    doi = "10.1038/35065675",
    number = "6825",
    urldate = "2025-08-08",
    journal = "Nature",
    year = "2001",
    pages = "242--250"
}

@article{spaepen_microscopic_1977,
    author = "Spaepen, F.",
    title = "A microscopic mechanism for steady state inhomogeneous flow in metallic glasses",
    volume = "25",
    copyright = "https://www.elsevier.com/tdm/userlicense/1.0/",
    issn = "00016160",
    doi = "10.1016/0001-6160(77)90232-2",
    number = "4",
    urldate = "2025-08-08",
    journal = "Acta Metal.",
    year = "1977",
    pages = "407--415"
}

@article{Henkelman2000NEB,
    author = "Henkelman, G. and Jónsson, H.",
    title = "Improved tangent estimate in the nudged elastic band method for finding minimum energy paths and saddle points",
    journal = "J. Chem. Phys.",
    volume = "113",
    number = "22",
    pages = "9978-9985",
    year = "2000",
    issn = "0021-9606",
    doi = "10.1063/1.1323224"
}

@article{Henkelman2000CI-NEB,
    author = "Henkelman, G. and Uberuaga, B. P. and Jónsson, H.",
    title = "A climbing image nudged elastic band method for finding saddle points and minimum energy paths",
    journal = "J. Chem. Phys.",
    volume = "113",
    number = "22",
    pages = "9901-9904",
    year = "2000",
    issn = "0021-9606",
    doi = "10.1063/1.1329672"
}

@article{Malandro1999JCP,
    author = "Malandro, D. L. and Lacks, D. J.",
    title = "Relationships of shear-induced changes in the potential energy landscape to the mechanical properties of ductile glasses",
    journal = "J. Chem. Phys.",
    volume = "110",
    number = "9",
    pages = "4593-4601",
    year = "1999",
    issn = "0021-9606",
    doi = "10.1063/1.478340"
}

@article{Maloney2006PRE,
    author = "Maloney, C. E. and Lema\^{\i}tre, A.",
    title = "Amorphous systems in athermal, quasistatic shear",
    journal = "Phys. Rev. E",
    volume = "74",
    issue = "1",
    pages = "016118",
    numpages = "22",
    year = "2006",
    publisher = "American Physical Society",
    doi = "10.1103/PhysRevE.74.016118"
}

@article{Rodney2011MSMSE,
    author = "Boioli, F. and Albaret, T. and Rodney, D.",
    title = "Shear transformation distribution and activation in glasses at the atomic scale",
    journal = "Phys. Rev. E",
    volume = "95",
    issue = "3",
    pages = "033005",
    numpages = "10",
    year = "2017",
    publisher = "American Physical Society",
    doi = "10.1103/PhysRevE.95.033005"
}

@article{Caroli_rate,
    author = "Lema\^{\i}tre, A. and Caroli, C.",
    title = "Rate-Dependent Avalanche Size in Athermally Sheared Amorphous Solids",
    journal = "Phys. Rev. Lett.",
    volume = "103",
    issue = "6",
    pages = "065501",
    numpages = "4",
    year = "2009",
    publisher = "American Physical Society",
    doi = "10.1103/PhysRevLett.103.065501"
}

@article{Edan_temp,
    author = "Karmakar, S. and Lerner, E. and Procaccia, I. and Zylberg, J.",
    title = "Statistical physics of elastoplastic steady states in amorphous solids: Finite temperatures and strain rates",
    journal = "Phys. Rev. E",
    volume = "82",
    issue = "3",
    pages = "031301",
    numpages = "11",
    year = "2010",
    publisher = "American Physical Society",
    doi = "10.1103/PhysRevE.82.031301"
}

@article{phase_diagram,
    author = "Korchinski, D. and Rottler, J.",
    title = "Dynamic phase diagram of plastically deformed amorphous solids at finite temperature",
    journal = "Phys. Rev. E",
    volume = "106",
    issue = "3",
    pages = "034103",
    numpages = "11",
    year = "2022",
    publisher = "American Physical Society",
    doi = "10.1103/PhysRevE.106.034103"
}

@misc{note1,
    note = "Note that a pre-released version of this potential was used to avoid a known problem with numerical instabilities affecting relaxations due to the explicit pairwise dispersion correction added to the published potential."
}

@article{Salerno_2012,
    author = "Salerno, K. Michael and Maloney, C. E. and Robbins, M. O.",
    title = "Avalanches in Strained Amorphous Solids: Does Inertia Destroy Critical Behavior?",
    volume = "109",
    doi = "10.1103/PhysRevLett.109.105703",
    abstractNote = "Simulations are used to determine the effect of inertia on athermal shear of amorphous two-dimensional solids. In the quasistatic limit, shear occurs through a series of rapid avalanches. The distribution of avalanches is analyzed using finite-size scaling with thousands to millions of disks. Inertia takes the system to a new underdamped universality class rather than driving the system away from criticality as previously thought. Scaling exponents are determined for the underdamped and overdamped limits and a critical damping that separates the two regimes. Systems are in the overdamped universality class even when most vibrational modes are underdamped.",
    number = "10",
    journal = "Phys. Rev. Lett.",
    publisher = "American Physical Society",
    year = "2012",
    pages = "105703"
}

@article{Moriel_2024,
    author = "Moriel, A. and Richard, D. and Lerner, E. and Bouchbinder, E.",
    title = "Elementary processes in dilatational plasticity of glasses",
    volume = "6",
    doi = "10.1103/PhysRevResearch.6.023167",
    abstractNote = "Materials typically fail under complex stress states, essentially involving dilatational (volumetric) components that eventually lead to material decohesion/separation. It is therefore important to understand dilatational irreversible deformation—i.e., dilatational plasticity—en route to failure. In the context of glasses, much focus has been given to shear (volume-preserving) plasticity, both in terms of the stress states considered and the corresponding material response. Here, using a recently developed methodology and extensive computer simulations, we shed basic light on the elementary processes mediating dilatational plasticity in glasses. We show that plastic instabilities, corresponding to singularities of the glass Hessian, generically feature both dilatational and shear irreversible strain components. The relative magnitude and statistics of the strain components depend both on the symmetry of the driving stress (e.g., shear versus hydrostatic tension) and on the cohesive (attractive) part of the interatomic interaction. We further show that the tensorial shear component of the plastic strain is generally nonplanar and also extract the characteristic volume of plastic instabilities. Elucidating the fundamental properties of the elementary micromechanical building blocks of plasticity in glasses sets the stage for addressing larger-scale, collective phenomena in dilatational plasticity such as topological changes in the form of cavitation and ductile-to-brittle transitions. As a first step in this direction, we show that the elastic moduli markedly soften during dilatational plastic deformation approaching cavitation.",
    number = "2",
    journal = "Phys. Rev. Res.",
    publisher = "American Physical Society",
    year = "2024",
    pages = "023167"
}

@article{Oyama_2024,
    author = "Oyama, N. and Kawasaki, T. and Kim, K. and Mizuno, H.",
    title = "Scale Separation of Shear-Induced Criticality in Glasses",
    volume = "132",
    doi = "10.1103/PhysRevLett.132.148201",
    abstractNote = "In a sheared steady state, glasses reach a nonequilibrium criticality called yielding criticality. We report that the qualitative nature of this nonequilibrium critical phenomenon depends on the details of the system and that responses and fluctuations are governed by different critical correlation lengths in specific situations. This scale separation of critical lengths arises when the screening of elastic propagation of mechanical signals is not negligible. We also discuss the determinant of the impact of screening effects from the viewpoint of the microscopic dissipation mechanism.",
    number = "14",
    journal = "Phys. Rev. Lett.",
    publisher = "American Physical Society",
    year = "2024",
    pages = "148201"
}

@article{Dattani_2022,
    author = "Dattani, U. A. and Karmakar, S. and Chaudhuri, P.",
    title = "Universal mechanical instabilities in the energy landscape of amorphous solids: Evidence from athermal quasistatic expansion",
    volume = "106",
    doi = "10.1103/PhysRevE.106.055004",
    abstractNote = "Using numerical simulations, we study the failure of an amorphous solid under athermal quasistatic expansion starting from a homogeneous high-density state. During the expansion process, plastic instabilities occur, manifested via sudden jumps in pressure and energy, with the largest event happening via cavitation leading to the material’s yielding. We demonstrate that all these plastic events are characterized by saddle-node bifurcation, during which the smallest nonzero eigenvalue of the Hessian matrix vanishes via a square-root singularity. We find that after yielding and prior to complete fracture, the statistics of pressure or energy jumps corresponding to the plastic events show subextensive system-size scaling, similar to the case of simple shear but with different exponents. Thus, overall, our paper reveals universal features in the fundamental characteristics during mechanical failure in amorphous solids under any quasistatic deformation protocol.",
    number = "5",
    journal = "Phys. Rev. E",
    publisher = "American Physical Society",
    year = "2022",
    pages = "055004"
}

@article{Duan_2023,
    author = "Duan, J. and Wang, Y. J. and Dai, L. H. and Jiang, M. Q.",
    title = "Elastic interactions of plastic events in strained amorphous solids before yield",
    volume = "7",
    doi = "10.1103/PhysRevMaterials.7.013601",
    abstractNote = "It has been widely accepted that the plastic events of amorphous solids after mechanical yield belong to a highly correlated avalanche state. However, whether the plastic events before yield are correlated or not is still unsettled, leaving their interactions largely unexplored. In this paper, by means of atomistic simulations, typical Cu50⁢Zr50 metallic glasses, as the model system, are sheared under athermal quasistatic limit to study these plastic events. The statistical analysis of both stress drops and waiting times reveals that plastic events before yield are in the correlated avalanche state and the interactions among them are mediated by the robust elasticity. The temporal correlation analysis of the nonaffine displacement fields further reveals that the elastic interactions are short-lived strong but long-standing weak, which results in the fractal morphology of potential energy landscape. By introducing vibrational modes to explore plastic events, we clearly exhibit the way how the elastic interactions organize the Eshelby-type shear transformations into avalanched plastic events. The correlation matrix, with its component being the dot product of the vibrational modes at different configurations, is defined to trace the evolution of vibrational modes during elastic deformation and across plastic events. Three reasons accounting for the robust elasticity are identified: (i) the limited destruction of plastic events on global elasticity, (ii) the persistent hard spots embedded in elastic matrix, and (iii) the self-recovery of elastic matrix during elastic deformation. Our results clarify the atomic-scale nature of both elastic deformation and plastic instabilities before yield in amorphous solids, providing fundamental information for the development of elastoplastic constitutive models.",
    number = "1",
    journal = "Phys. Rev. Mater.",
    publisher = "American Physical Society",
    year = "2023",
    pages = "013601"
}

@ARTICLE{Liu2016-tc,
    author = "Liu, C. and Ferrero, E. E and Puosi, F. and Barrat, J. and Martens, K.",
    title = "Driving Rate Dependence of Avalanche Statistics and Shapes at the Yielding Transition",
    journal = "Phys. Rev. Lett.",
    publisher = "American Physical Society",
    volume = "116",
    number = "6",
    pages = "065501",
    year = "2016",
    doi = "10.1103/PhysRevLett.116.065501"
}

@software{Buze_LACT_LAMMPS_Continuation,
author = {Buze, Maciej and Birks, Fraser},
title = {{LACT (LAMMPS Continuation Techniques)}},
note = {https://github.com/mbuze/LACT}
}

@article{Salman_2011, title={Minimal Integer Automaton behind Crystal Plasticity}, volume={106}, DOI={10.1103/PhysRevLett.106.175503}, number={17}, journal={Phys. Rev. Lett.}, publisher={American Physical Society}, author={Salman, Oğuz Umut and Truskinovsky, Lev}, year={2011}, month=apr, pages={175503} }

@article{Baggio_2023, title={Inelastic rotations and pseudoturbulent plastic avalanches in crystals}, volume={107}, DOI={10.1103/PhysRevE.107.025004}, number={2}, journal={Phys. Rev. E}, publisher={American Physical Society}, author={Baggio, R. and Salman, O. U. and Truskinovsky, L.}, year={2023}, month=feb, pages={025004} }

@article{Kunze_2014, title={Wear, Plasticity, and Rehybridization in Tetrahedral Amorphous Carbon}, volume={53}, ISSN={1573-2711}, DOI={10.1007/s11249-013-0250-7}, number={1}, journal={Tribology Letters}, author={Kunze, Tim and Posselt, Matthias and Gemming, Sibylle and Seifert, Gotthard and Konicek, Andrew R. and Carpick, Robert W. and Pastewka, Lars and Moseler, Michael}, year={2014}, month=jan, pages={119–126}, language={en} }

@article{quickmin,
    author = {Sheppard, Daniel and Terrell, Rye and Henkelman, Graeme},
    title = {Optimization methods for finding minimum energy paths},
    journal = {The Journal of Chemical Physics},
    volume = {128},
    number = {13},
    pages = {134106},
    year = {2008},
    month = {04},
    issn = {0021-9606},
    doi = {10.1063/1.2841941},
    url = {https://doi.org/10.1063/1.2841941},
}

@article{Heun,
    author = {Sammüller, Florian and Schmidt, Matthias},
    title = {Adaptive Brownian Dynamics},
    journal = {The Journal of Chemical Physics},
    volume = {155},
    number = {13},
    pages = {134107},
    year = {2021},
    month = {10},
    doi = {10.1063/5.0062396},
}

@dataset{Birks2026ArclengthData,
  author       = {Birks, Fraser and Ghanem, Imad and Pastewka, Lars and Kermode, James and Buze, Mihai},
  title        = {Supporting Code and Data for {``Resolving Structural Avalanches in Amorphous Carbon with Arclength Continuation''}},
  year         = {2026},
  version      = {v0.1},
  publisher    = {Zenodo},
  doi          = {10.5281/zenodo.18393712},
  url          = {https://doi.org/10.5281/zenodo.18393712},
}

@article{Pastewka2013,
  author  = {Pastewka, Lars and Klemenz, Andreas and Gumbsch, Peter and Moseler, Michael},
  title   = {Screened empirical bond-order potentials for {Si-C}},
  journal = {Physical Review B},
  volume  = {87},
  number  = {20},
  pages   = {205410},
  year    = {2013},
  doi     = {10.1103/PhysRevB.87.205410}
}

@article{Baskes_1994,
    doi = {10.1088/0965-0393/2/3A/006},
    url = {https://doi.org/10.1088/0965-0393/2/3A/006},
    year = {1994},
    month = {may},
    publisher = {},
    volume = {2},
    number = {3A},
    pages = {505},
    author = {M I Baskes and J E Angelo and C L Bisson},
    title = {Atomistic calculations of composite interfaces},
    journal = {Modelling and Simulation in Materials Science and Engineering}
}

@misc{SupplementalA,  year = {2026}, note = "See Supplemental Material A for an investigation into the use of nudged elastic band simulations for resolving avalanches."}

@misc{SupplementalB,  year = {2026},
  note = "See Supplemental Material B for detailed descriptions of the arclength continuation algorithms used in this study."
}

@misc{SupplementalC,  year = {2026},
  note = "See Supplemental Material C for an overview of the avalanche sizes investigated in this study and details on method limitations."
}

@misc{SupplementalD,  year = {2026},
  note = "See Supplemental Material D for an investigation into the reproducibility of stress strain curves with both the AC and AQS methods."
}

@misc{SupplementalE,  year = {2026},
  note = "See Supplemental Material E for a comparison between the ordering of plastic events in avalanches derived using arclength continuation and damped dynamics"
}

\end{document}


\section*{Supplementary material C\\[10pt] Separability of Avalanches}
$\,$
\section{Investigating separability}
A central issue in the study of avalanches in amorphous carbon is their separability -- that is, the extent to which they can be decomposed into transitions between intermediate basins occurring at strains below the avalanche onset.
To investigate this, we applied our numerical continuation algorithm to explore the energy landscapes of the first 51 plastic events across three different 4096 atom amorphous carbon systems (a total of 153 events). For each event, we recorded (i) the number of bonds involved in the avalanche and (ii) the `degree of separability' (how many intermediate states we could recover at lower strains).

\begin{figure*}
    \centering
    \includegraphics[width=0.9\textwidth]{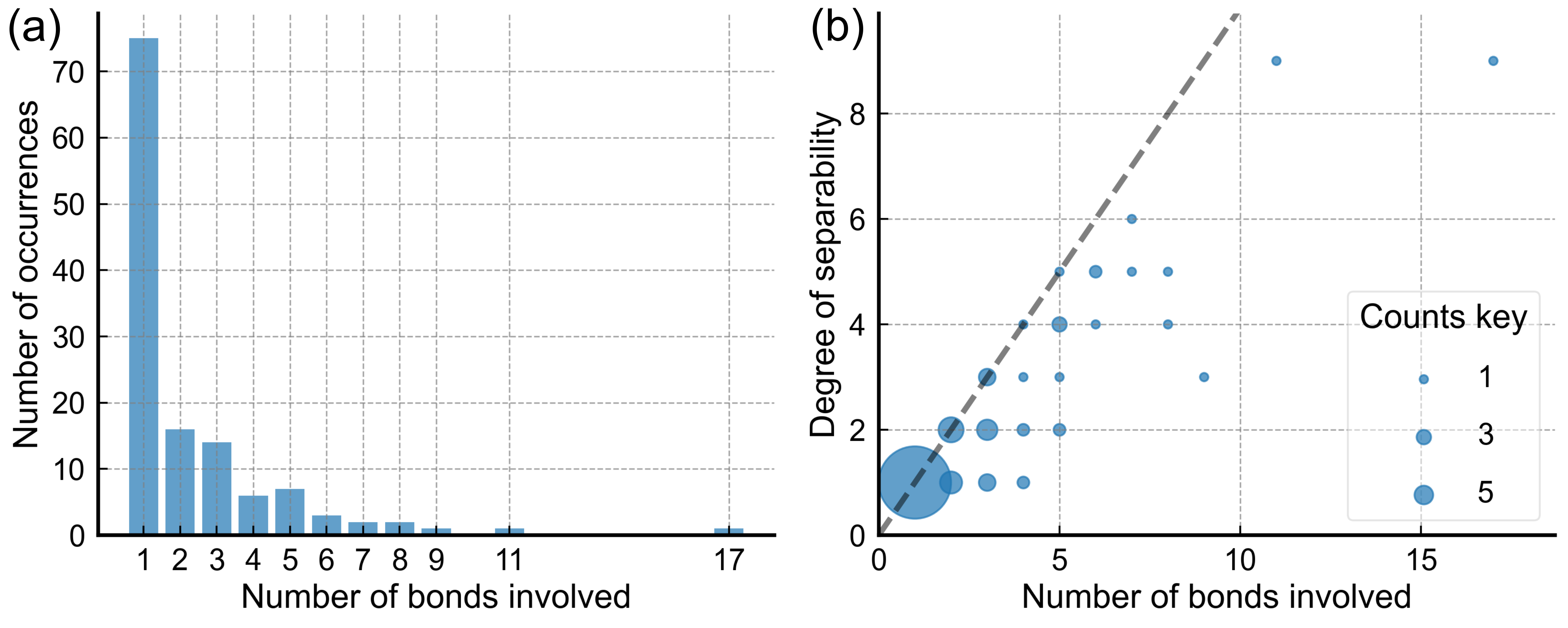}
    \caption{A demonstration of the size and separability of the first 51 plastic events across three amorphous carbon structures that were successfully explored by the algorithm. (a) The occurrence of events by number of bonds. More than half of the events involve just one bond. (b) Degree of separability of different plastic events. The majority of events involving multiple bonds are at least partially separable. The dashed line represents `perfect separability', with points on that line corresponding to avalanches that have been perfectly separated into chains of single bond events.}
    \label{fig:bond_num_si}
\end{figure*}

This information is displayed for all events successfully explored in Fig.~\ref{fig:bond_num_si}. The distribution of the number of bonds involved in plastic events is shown in Fig.~\ref{fig:bond_num_si}a. It can be seen that the majority of plastic events involve only a single bond, with larger plastic events becoming increasingly less common. The separability of plastic events is shown as a scatter plot in Fig.~\ref{fig:bond_num_si}b. The dashed line represents `perfect separability' -- a perfectly separable avalanche involving N bonds can be split into N discrete transitions at strains below the onset. We found that the majority of plastic events involving multiple bonds were separable to some extent, but relatively few were perfectly separable. 

We found three explanations for this. First, it was common for the same atom to be involved in the breaking of one bond and the forming of another during a single plastic event. Our criteria classified such an event as involving two bonds, but energetically this transition behaves like an inseparable single-bond event. Second, on occasion, the bond detection algorithm mistakenly classified a bond as breaking or forming during an avalanche when in reality it only stretched or compressed slightly - erroneously inflating the number of bonds involved. We minimised these errors by applying a finely tuned two-cutoff bond classification algorithm (described in the next section). Finally (and most commonly), the single bond events were so correlated that they could not be disentangled at lower strains.

It is important to make a distinction between the highly correlated chains of single-bond events seen here and the collective motion of many atoms (like one sees, for example, in Lennard-Jones metallic glass systems). Crucially, in amorphous carbon, even in the case of an inseparable correlated chain of single bond events, the initial saddle point is index-1 and involves only a single bond.

\subsection{Bond-identification procedure}

Because the interpretation of avalanche size depends directly on which pairs of atoms are classified as bonded, it is important to clarify that we do not use a single first-neighbour distance cutoff. In amorphous carbon, there is no unique radial threshold that cleanly separates bonded from non-bonded C--C pairs across all local environments. There exists an intermediate-distance regime (between the first and second neighbor shells in the radial distribution function) where C–C pairs can be either bonded or non-bonded depending on their local coordination geometry.

Instead, bonds were identified using a two-threshold plus local-screening procedure based on the screening formalism used in modified embedded-atom method and bond-order potentials~\cite{Baskes_1994,Pastewka2013}. For a candidate pair of atoms $i$ and $j$ separated by a distance $r_{ij}$, we apply the following rules:
\begin{enumerate}
    \item if $r_{ij} < r_{\mathrm{in}} = 1.823$~\AA, the pair is classified as bonded;
    \item if $r_{ij} > r_{\mathrm{out}} = 2.195$~\AA, the pair is classified as non-bonded;
    \item if $r_{\mathrm{in}} \le r_{ij} \le r_{\mathrm{out}}$, the classification is determined from the local bonding environment.
\end{enumerate}

For pairs in this intermediate regime, we identify all atoms $k$ that are neighbours of both $i$ and $j$, and evaluate the screening quantity
\begin{equation}
    C_{ijk}
    =
    \frac{
        2(x_{ik}+x_{jk})-(x_{ik}-x_{jk})^{2}-1
    }{
        1-(x_{ik}-x_{jk})^{2}
    },
\end{equation}
where
\begin{equation}
    x_{ik} = \left(\frac{r_{ik}}{r_{ij}}\right)^{2},
    \qquad
    x_{jk} = \left(\frac{r_{jk}}{r_{ij}}\right)^{2}.
\end{equation}
If at least one common neighbour satisfies
\begin{equation}
    0 \le C_{ijk} \le 3,
\end{equation}
then the pair $i$--$j$ is treated as screened and therefore non-bonded. Otherwise, it is retained as a bond. In this way, ambiguous intermediate-distance pairs are resolved using their immediate local topology rather than distance alone.

This procedure is substantially more robust than a single-distance cutoff for amorphous carbon, and avoids many spurious bond-formation and bond-breaking detections. However, it is not faultless, and some residual misclassification remains possible. Developing a more accurate classifier based on machine-learned local-environment descriptors would be an interesting direction for future work.

\section{Method limitations and improvements}

\begin{figure*}
    \centering
    \includegraphics[width=0.9\textwidth]{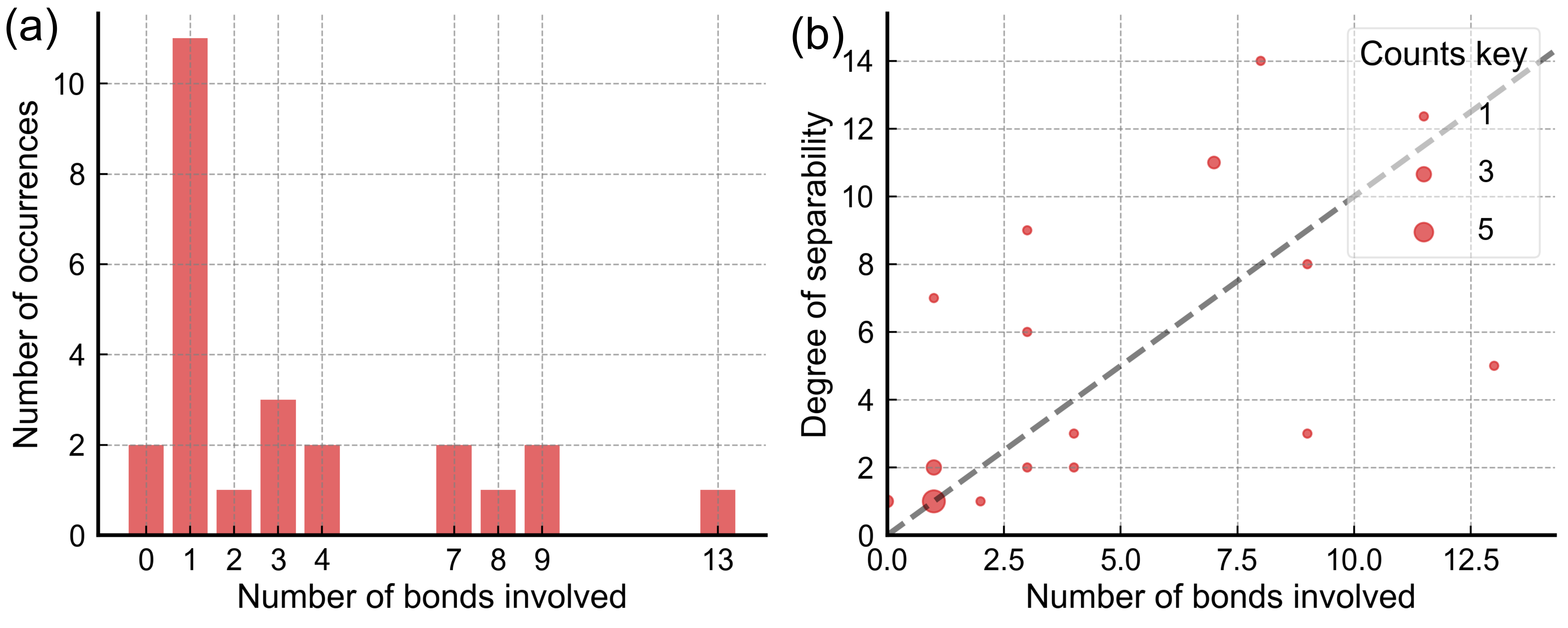}
    \caption{A demonstration of the size and separability of the first 51 plastic events across three amorphous carbon structures that the algorithm failed to correctly explore. (a) The occurrence of events by number of bonds. Note that the the algorithm disproportionately fails on larger avalanches in the plastic flow regime. (b) Degree of separability of different plastic events. Many events are 'more than perfectly separable', indicating that other bonds broke during the exploration which were not part of the initial avalanche. Most failures are due to the relaxation algorithm overcoming small barriers to trigger events that were not part of the initial avalanche.}
    \label{fig:bond_num_si_2}
\end{figure*}

Of the 153 events analyzed, there were 25 instances where the algorithm failed during the exploration of an avalanche. The statistics of these failures can be seen in Fig.~\ref{fig:bond_num_si_2}. For the most part, these failures can be attributed to the slight unpredictability of the relaxation stage of the algorithm (where relaxations are initiated from uniformly sampled saddle points on an unstable segment). For the algorithm to perform perfectly, it is essential that single-bond events only trigger if they have a barrier of zero. However, in practice, when there are many small barriers in the system (as is common in the plastic flow regime), it is common for small ($\sim \mathrm{meV}$) barriers to be overcome during a relaxation. If this happens, neighboring local minima can involve very different subsets of bonds in the full avalanche causing the exploration to fail.

One possible way of overcoming this would be to replace the relaxation with a continuation using a more suitable (and localized) continuation coordinate than the strain on the system. Selecting such a coordinate for the case of a single-bond is simple -- one can just use a force applied to the bond (as is done in Algorithm 2 in Supplementary Material B). Selecting a suitable coordinate for multi-bond avalanches however is a non-trivial endeavor, and is therefore left as an interesting direction for future work.

\bibliographystyle{apsrev4-2} 
\bibliography{refs.bib} 


\section*{Supplementary material B\\[10pt] Tailored Hessian-free numerical continuation for atomistic modelling with machine learned interatomic potentials}
$\,$

\subsection*{B1 Periodic MLIP energy without restricting positions to the simulation cell}

Let $A \in \mathbb{R}^{3 \times 3}$ be a matrix specifying a simulation cell $\Omega(A) := \{ A \bm\xi:0\leq\xi_i<1\}$. Further, let $\mathcal{L}(A)=\{A\mathbf n:\mathbf n\in\mathbb Z^3\}$ denote the Bravais lattice spawned by $A$. Denote the atomic coordinates of the $N$ atoms being modelled by
\[
\mathbf R=(\mathbf r_1,\dots,\mathbf r_N)\in\mathbb R^{3N},
\]
where, to make the computation of tangents in the numerical continuation scheme intuitive, \emph{each} $\mathbf r_i$ is allowed to be any point in $\mathbb R^3$,
i.e. we do \emph{not} impose $\mathbf r_i\in\Omega(A)$.

Fix a finite interaction cutoff $r_c>0$. Define the
\emph{neighbour-image set} of the $i$th atom by 
\[
\mathcal N_i(\mathbf R,A)
:=\{(j,\mathbf n)\in\{1,\dots,N\}\times\mathbb Z^3 \;:\;
\|\mathbf r_j + A\mathbf n - \mathbf r_i\| < r_c\}.
\]
Each element $(j,\mathbf n)$ denotes the image of the $j$th atom
translated by $A\mathbf n$. Form the multiset of relative vectors
\[
\mathcal R_i(\mathbf R,A)
:=\{\; \mathbf r_j + A\mathbf n - \mathbf r_i \;:\; (j,\mathbf n)\in\mathcal N_i(\mathbf r,A)\;\}.
\]

Let $\mathcal{E}_i$ be a general site energy mapping a finite multiset of relative vectors to $\mathbb R$ (thus encompassing general many–body descriptors in MLIPs). Then define the periodic energy
\begin{equation}\label{eqn:E_r_A}
	\mathcal{E}\,\colon\,\mathbb{R}^{3N} \times \mathbb{R} \to \mathbb{R},\quad \mathcal{E}(\mathbf R,\lambda)\;=\;\sum_{i=1}^N U_i\!\big(\mathcal R_i(\mathbf r,A_{\lambda}) \big), 
\end{equation}
where $A_{\lambda} := T_{\lambda}A_0$, with $A_0$ the matrix defining the default simulation cell and $T_{\lambda}$ a deformation gradient corresponding to a simple shear $\lambda = \varepsilon_{xy}$. To be precise, 
	\[
	T_{\lambda} = \begin{pmatrix}
			1 & \lambda & 0 \\ 0 & 1 & 0 \\ 0 & 0 & 1
		\end{pmatrix}.
	\]

The forces acting on the $N$ atoms are defined as
\[
\mathbf{F}\,\colon\, \mathbb{R}^{3N} \times \mathbb{R} \to \mathbb{R}^{3N}, \quad \mathbf{F}(\mathbf{R},\lambda) := \nabla_{\mathbf{R}}\mathcal{E}(\mathbf{R},\lambda).
\]

\subsection*{B2 Hessian-free numerical continuation for MLIP systems}

Numerical continuation is a set of tools to trace continuous paths of the form
\begin{equation}\label{eqn:sol-path}
	\{ (\mathbf R(s),\lambda(s)) \in \mathbb{R}^{3N+1} \,\mid\, s \in (0,1),\quad \mathbf{F}(\mathbf{R}(s),\lambda(s)) = \mathbf{0}\}.
\end{equation}
For a sufficiently locally smooth atomistic energy $\mathcal E$ of the form given by \eqref{eqn:E_r_A}, the existence of such a path of equilibrium configurations in the vicinity of an already known pair $(\mathbf{R}_k,\lambda_k)$ is guaranteed by the Implicit Function Theorem \cite{allgower2003introduction}. The parameter $s$ is interpreted as arclength along the solution branch. 

To find a nearby pair on the solution branch \eqref{eqn:sol-path}, one first computes the tangent vector
\begin{equation}\label{eqn:tangent}
	(\dot{\mathbf{R}}_k, \dot \lambda_k) := \frac{\mathrm{d}}{\mathrm{d}s}(\mathbf{R}_k(s),\lambda_k(s)),
\end{equation}
and then solves an extended system of equations $\mathbf{G}(\mathbf{R}_{k+1},\lambda_{k+1}) = 0$, where
\begin{equation}\label{eqn:extended_system}
	\mathbf{G}\colon \mathbb{R}^{3N+1} \to \mathbb{R}^{3N+1},\quad 
	\mathbf{G}(\mathbf{R},\lambda) := 
	\begin{pmatrix}
		\mathbf{F}(\mathbf{R},\lambda)\\
		(\mathbf{R} - \mathbf{R}_{k})^{\top}\dot{\mathbf{R}}_k + (\lambda - \lambda_{k}) \dot \lambda_k - \delta s
	\end{pmatrix},
\end{equation}
starting from the initial guess $(\mathbf{R}_k + \delta s \dot{\mathbf{R}}_k, \lambda_k + \delta s \dot \lambda_k)$. 
It is well known \cite{allgower2003introduction} that this extended system remains regular even at points where the Jacobian of $\mathbf F$, denoted $\nabla_{\mathbf{R}}\mathbf{F}$ (the Hessian of the atomistic energy $\nabla^2_{\mathbf{R}}\mathcal E$), is singular. Moreover, for sufficiently small $\delta s > 0$, it can be rigorously shown that the new solution $(\mathbf{R}_{k+1},\lambda_{k+1})$ remains on the same branch \eqref{eqn:sol-path} \cite{allgower2003introduction}.

The robustness of the method relies on accurate tangent computation. In principle, the tangent \eqref{eqn:tangent} is obtained by differentiating the condition in \eqref{eqn:sol-path} with respect to $s$, which requires access to $\nabla_{\mathbf{R}}\mathbf{F} = \nabla^2_{\mathbf{R}}\mathcal{E}$. This is prohibitively expensive for modern MLIPs, so we instead approximate the tangent using finite differences between previous solutions. Similarly, to avoid computing the full Hessian when solving \eqref{eqn:extended_system}, we use a Krylov-subspace approach, which only requires the ability to compute matrix-vector products with $\nabla_{\mathbf R} \mathbf{F}$, which themselves can be approximated by finite differences. Related ideas have recently been explored in \cite{maliyov2025exploring}. Our continuation code, LACT \cite{Buze_LACT_LAMMPS_Continuation} is a Python-based wrapper around the molecular dynamics code LAMMPS \cite{LAMMPS}.

\subsection*{B3 Ensuring relevance of computed branches}

The energy landscape associated with $\mathcal E$ is highly complex, and
not every continuation solution branch of the form \eqref{eqn:sol-path} obtained with base continuation routine described in A2 is physically relevant.
To isolate the meaningful branches, we employ the automated procedure detailed in Algorithm~\ref{alg:continuation1}. The energy plots shown in the main text are based directly on outputs of this scheme.

\begin{algorithm}[h!]
	\caption{Recursive continuation with avalanche dissection}
	\label{alg:continuation1}
	\KwIn{Initial minimum $\mathbf R_0$ at parameter $\lambda_0$}
	\KwOut{Set of physically relevant continuation solution branches}
	\BlankLine
	
	\textbf{Stage 1: Accelerated quasi-static loading}\\
	Run continuation algorithm until increment $\Delta \lambda$ changes sign (bifurcation point). \\
	Perform quasi-static loading to trigger avalanche; 
	record bonds involved \\
	
	\textbf{Stage 2: Saddle-point tracing}\\
    Use Algorithm \ref{alg:continuation2} to traverse sharp bifurcation point onto the saddle. \\
	Restart continuation and run on saddle until $\Delta \lambda$ changes sign. \\

    \textbf{Stage 3: Restart or termination} \\
	\For{a uniformly sampled subset of saddle points $(\mathbf{R},\lambda)$:}{Take current saddle $\mathbf R_k$ and relax, obtaining local minima $\mathbf R_k^0$. Perturb $\mathbf R_k$ by $\delta \frac{\mathbf{R}_k - \mathbf{R}_k^0}{|\mathbf{R}_k - \mathbf{R}_k^0|}$ and relax again, obtaining a second local minima $\mathbf{R}_k^1$.}
	\ForEach{newly identified minima pair}{
		\If{the structural change between them only involves a previously unexplored subset of avalanche bonds}{
			Record as a candidate restart position.
		}
		\Else{
			Disregard as irrelevant.
		}
	}
    \ForEach{candidate restart position}{
        \If{candidate has highest $\lambda$ in a chain of at least two candidate restarts which each have the same subset of avalanche bonds}{
            Keep candidate
        }
        \Else{
            Discard candidate
        }
    }
    \If{At least one candidate position remaining}{
        \If{If only remaining candidate involves all remaining bonds in the avalanche}{
            Terminate, marking avalanche as successfully explored.
        }
        \Else{Restart algorithm from the highest $\lambda$ candidate showing the smallest subset of avalanche bonds.}
    }
	\Else{Terminate, marking avalanche as not fully explored.}
\end{algorithm}

\subsection*{B4 Overcoming numerical stability issues with traversing bifurcation points}

In practice, some bifurcation points involve such a sharp turn with respect to $\lambda$ that a miniscule $\delta s$ is required to traverse it. This in turn leads to numerical stability issues when computing tangents with finite difference approximations, often leading to a solution branch artificially turning on itself. To avoid this, close to a bifurcation point, we switch to a force-based continuation scheme detailed in Algorithm~\ref{alg:continuation2}.
\begin{algorithm}[h!]
	\caption{Force-assisted continuation through bifurcation points}
	\label{alg:continuation2}
	\KwIn{Initial minimum $\mathbf R_0$ at parameter $\lambda_0$}
	\KwOut{Set of physically relevant continuation solution branches}
	\BlankLine
	
	\textbf{Stage 1: Accelerated quasi-static loading}\\
	Run continuation until the increment $\Delta \lambda$ changes sign (bifurcation point).\\
	Step back and perform quasi-static loading to trigger avalanche; 
	record the set of bonds $\{(\mathbf r_i,\mathbf r_j)\}$ involved. \\
	\textbf{Stage 2: Force-based continuation}\\
	Roll back five continuation steps. Among avalanche bonds, select $(\mathbf r_i,\mathbf r_j)$ with the largest $\tfrac{\mathrm d}{\mathrm d\lambda}|\mathbf r_i-\mathbf r_j|$ (approximated by finite differences).\\
	Freeze $\lambda$ and define an auxiliary force field 
	$\mathbf F^{\mu}(\mathbf R) = (\mathbf f^{\mu}_1,\dots,\mathbf f^{\mu}_N)$ with 
	\[
	\mathbf f^{\mu}_i = \mu \frac{\mathbf r_i-\mathbf r_j}{|\mathbf r_i-\mathbf r_j|}, 
	\quad 
	\mathbf f^{\mu}_j = -\mathbf f^{\mu}_i,
	\quad 
	\mathbf f^{\mu}_k = 0 \;\; (k \neq i,j),
	\]
	where $\mathbf F^{\mu}$ depends on current configuration $\mathbf R$.\\
	Initialise with $\mu=0$ and run continuation as in (A2), replacing $\mathbf F$ by $\mathbf F+\mathbf F^{\mu}$ and treating $\mu$ as the continuation parameter.\\
	
	\textbf{Stage 3: Return to standard continuation}\\
	Terminate the force-based routine once $\Delta \mu$ changes sign and subsequently $\mu$ returns to zero. The corresponding configuration is guaranteed to be a saddle point. \\
	\BlankLine
\end{algorithm}

\bibliographystyle{apsrev4-2}
\bibliography{refs.bib} 


\section*{Supplementary material A\\[10pt] Non-Uniqueness of Minimum Energy Paths\\ in Nudged Elastic Band Simulations}
$\,$

In amorphous solids under athermal quasi-static shear, a plastic event occurs when a local minimum of the potential-energy surface (PES) loses stability through a saddle-node (fold) bifurcation, in which the minimum collides with a nearby saddle, triggering a transition into a neighboring basin along the minimum-energy path (MEP)~\cite{Malandro1999JCP,Maloney2006PRE}.
To compute the MEP associated with a given plastic event, we use the Nudged Elastic Band (NEB) method~\cite{Henkelman2000NEB,Henkelman2000CI-NEB} following the procedure of Rodney and co-workers~\cite{Rodney2011MSMSE}.
Immediately after an instability is detected, the shear direction is reversed to recover two stable configurations at the same applied strain—just before and just after the event—which serve as the fixed endpoints for the subsequent NEB calculation.
Here we show that, for an avalanche comprising a sequence of bond transformations, the MEP between the same endpoints is not unique.
The NEB solution is highly sensitive to how the initial images are constructed, and straightforward linear interpolation does not necessarily recover the dynamically realized pathway at the instability.

\begin{figure*}
    \centering
    \includegraphics[width=\textwidth]{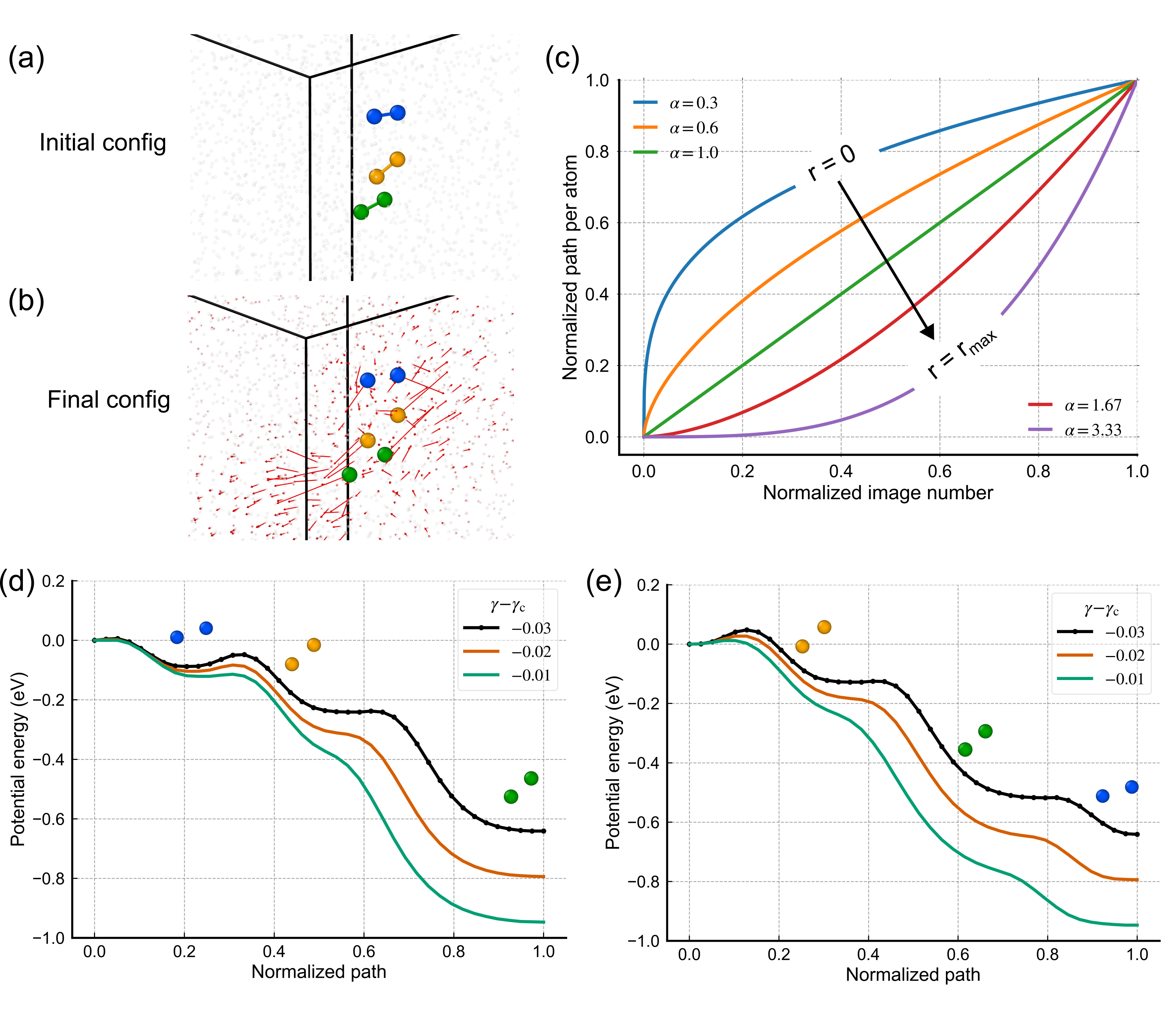}
    \caption{Dependence of the MEP on the initial path.
    (a,b) Initial and final configurations bracketing the avalanche at an offset of $\gamma-\gamma_{c}=-0.03$. Atoms participating in the avalanche are highlighted in distinct colors, while all other atoms, which respond elastically, are rendered small and semi-transparent for clarity. The displacement field (red arrows) between the initial and final configurations is overlaid on panel (b).
    (c) Per-atom interpolation used to construct the initial images between the NEB endpoints: a power-law progression with exponent $\alpha$ that depends on the atom’s distance $r$ from an arbitrarily chosen center ($r=0$).
    (d) MEPs obtained from NEB calculations with the initial images biased toward the blue bond at different values of strain offset $\gamma-\gamma_{c}$.
    (e) MEPs obtained when the initial images are biased toward the orange or green bond; both initializations converge to the same final paths.
    In both (d) and (e), atoms undergoing a plastic transformation are placed at their corresponding locations along the transition segments of the MEP.}
    \label{fig:neb_si}
\end{figure*}

We analyze an avalanche comprising three bond-breaking events.
At NEB endpoints, we select the configurations immediately before and after the avalanche at strain offsets $\gamma-\gamma_\mathrm{c}=-0.03$, $-0.02$, and $-0.01$, where $\gamma$ denotes the imposed shear strain and $\gamma_\mathrm{c}$ is the critical strain at which the instability occurs.
We show zoomed views of the atomic configurations at $\gamma-\gamma_\mathrm{c}=-0.03$ in Fig.~\ref{fig:neb_si}a and b, with the displacement field between them overlaid on panel~b. 
The atoms involved in the $3$-bond transformations are highlighted in distinct colors. 
Instead of linearly interpolating the displacement field between the two configurations, we adopt a nonlinear interpolation scheme that biases the initial MEP guess toward one of the highlighted bonds. 
We construct $N$ initial images by prescribing a per-atom power-law progression of the displacement along the path 
\begin{equation}
\bm{D}_j^i = \bigl(\bm{R}_j^N - \bm{R}_j^1\bigr)\left(\frac{i}{N}\right)^{\alpha_j},
\label{power-law}
\end{equation}
where $\bm{D}_j^i$ is the displacement of atom $j$ at image $i$ ($i=1,...,N$).
$\bm{R}_j^1$ and $\bm{R}_j^N$ are the positions of atom $j$ in the endpoint configurations. 
The power-law exponent $\alpha_j$ is a per-atom quantity that skews the displacement toward the beginning of the path ($\alpha_j<1$) or toward the end ($\alpha_j>1$). 
$\alpha$ attains a minimum value $\alpha_\text{min}$ at an arbitrarily chosen center ($r=0$) and a maximum value $\alpha_\text{max}$ at $r=L\sqrt{3}/2$, where $L$ is the simulation-box length. 
This distance equals half the body diagonal and is the maximum minimum-image distance in a cubic periodic cell. 
For $r \in [0,L\sqrt{3}/2]$, $\alpha$ varies linearly as %
\begin{equation}
\alpha = \alpha_{\text{min}} + \bigl(\alpha_{\text{max}} - \alpha_{\text{min}}\bigr) \frac{2r}{L\sqrt{3}}.
\end{equation}

Figure~\ref{fig:neb_si}c summarizes how the per-atom interpolation profile depends on the exponent $\alpha$.
For the avalanche analyzed here, we choose $\alpha_{\min}=0.1$ and $\alpha_{\max}=10$.
Using this scheme, we construct three distinct initial paths by centering the interpolation ($r=0$) on each of the bonds involved in the avalanche.
An NEB calculation is then performed for each path using the damped-dynamics optimizer quickmin~\cite{quickmin}, with convergence declared when the 2-norm of the global force vector falls below $10^{-4}~\text{eV}\,\text{\AA}^{-1}$.

In Fig.~\ref{fig:neb_si}d, we present the MEPs obtained by biasing the initial images toward the blue bond.
The dotted black MEP (at $\gamma-\gamma_{c}=-0.03$) consists of a sequence of distinct energy minima separated by finite barriers.
Each transition between consecutive minima corresponds to the breaking of a single bond, which is highlighted along the transition segments of the MEP.
As the critical strain is approached, some barriers vanish; however, the bonds undergoing transformation remain well separated along the path.
Biasing the initial images toward the orange or green bond leads to convergence onto the same final MEPs, in which the avalanche is initiated by breaking the orange bond, as shown in Fig.~\ref{fig:neb_si}e.
For the particular avalanche considered here, we do not observe any converged MEP in which the green bond acts as the initiating event.
Nevertheless, for each investigated strain offset, the NEB calculations converge to two distinct MEPs depending on the initialization, indicating that multiple competing pathways exist and that the converged MEP is initialization-dependent.
One could, in principle, continue approaching the critical strain and determine the physically relevant pathway by identifying which barrier drops to zero first. However, sufficiently close to the instability, the barriers become very shallow, approaching saddle points, and NEB calculations no longer converge in a robust manner.

\bibliographystyle{apsrev4-2} 
\bibliography{refs.bib} 


\section*{Supplementary material E\\[10pt] Causality chain of events in avalanches}
$\,$

Our analysis in the main manuscript demonstrates that plastic avalanches in a-C networks possess a \emph{latent internal structure}, manifested as a hierarchy of energy barriers revealed by numerical continuation, each associated with the transformation of an individual bond.
This hierarchy naturally introduces a notion of causality within a single avalanche, whereby the activation of one bond-level instability facilitates subsequent transformations.
To explicitly probe this causal sequence, we study the real-time dynamical evolution of avalanches.
The system dynamics are modeled using overdamped equations of motion, appropriate for the strongly dissipative, athermal regime relevant to quasi-static deformation.
The atomic positions $\mathbf{r}(t)$ evolve according to
\begin{equation}
    \dot{\mathbf{r}}(t)
    =
    \frac{1}{\zeta}\,
    \mathbf{F}\!\left(\mathbf{r}(t),t\right)
    \, ,
    \label{eq:overdamped}
\end{equation}
where $\mathbf{F}$ denotes the total interatomic force and $\zeta$ is the damping coefficient controlling the relaxation rate.
Equation~\eqref{eq:overdamped} is integrated numerically using the Heun method~\cite{Heun}, a second-order explicit predictor--corrector scheme:
\begin{equation}
\begin{aligned}
    \text{(predictor)}\quad
    \tilde{\mathbf{r}}_{n+1}
    &=
    \mathbf{r}_n
    +
    \frac{\Delta t}{\zeta}\,
    \mathbf{F}\!\left(\mathbf{r}_n\right),
    \\
    \text{(corrector)}\quad
    \mathbf{r}_{n+1}
    &=
    \mathbf{r}_n
    +
    \frac{\Delta t}{2\zeta}
    \left[
        \mathbf{F}\!\left(\mathbf{r}_n\right)
        +
        \mathbf{F}\!\left(\tilde{\mathbf{r}}_{n+1}\right)
    \right].
\end{aligned}
\label{eq:heun}
\end{equation}
where $\Delta t$ denotes the integration timestep.
To initiate an avalanche, the system is first driven arbitrarily close to the critical strain at which the instability occurs.
A small additional shear strain increment of $\Delta\gamma = 10^{-5}$ is then applied to trigger the event.
Following this perturbation, the system is allowed to evolve dynamically according to Eq.~\eqref{eq:heun} with a damping coefficient $\zeta = 0.2~\mathrm{eV \cdot ps \cdot \AA^{-2}}$ and a timestep $\Delta t = 10^{-4}~\mathrm{ps}$.

\begin{figure}
    \centering
    \includegraphics[width=0.8\textwidth]{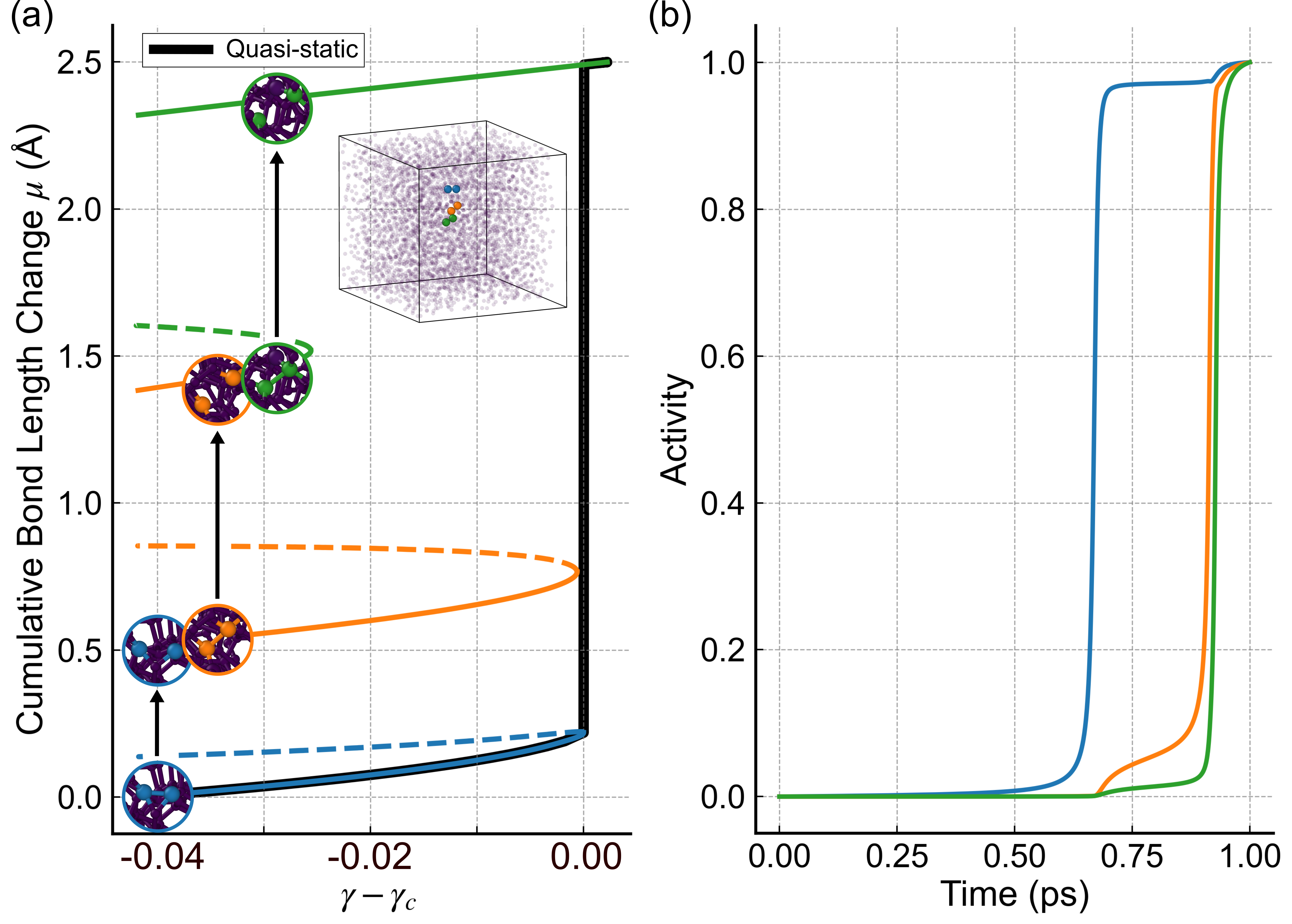}
    \caption{Chain of events within an avalanche involving three bond transformations. 
            (a) Hierarchy of bond-level instabilities obtained from numerical continuation, ordered from bottom to top. 
            (b) Corresponding overdamped dynamical evolution of the same bonds during the avalanche. 
            Colors are consistent between panels.}
    \label{fig_overdamped_small}
\end{figure}

\begin{figure}
    \centering
    \includegraphics[width=0.8\textwidth]{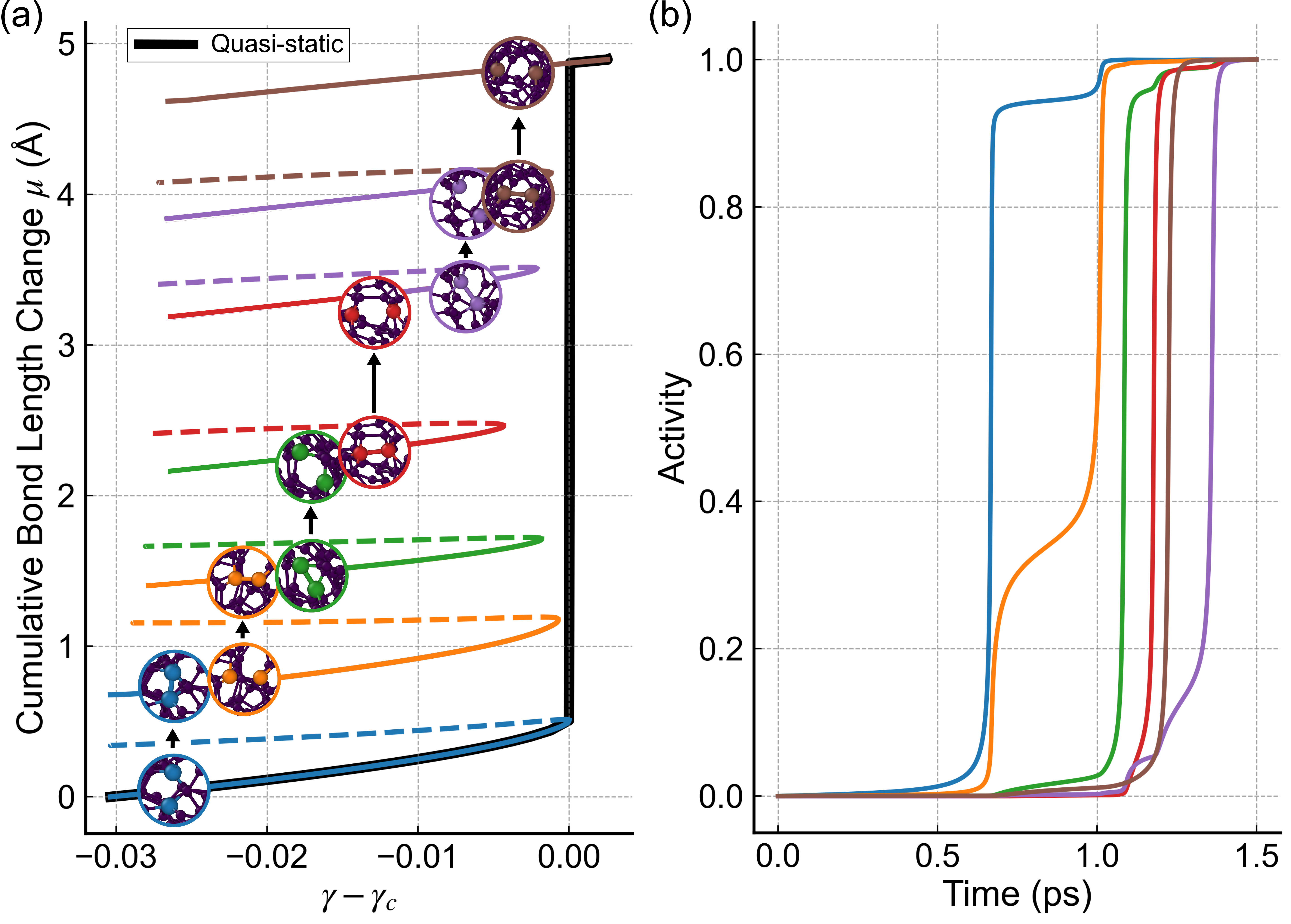}
    \caption{Chain of events within an avalanche involving six bond transformations. 
            (a) Hierarchy of bond-level instabilities obtained from numerical continuation, ordered from bottom to top. 
            (b) Corresponding overdamped dynamical evolution of the same bonds during the avalanche. 
            Colors are consistent between panels.}
    \label{fig_overdamped_big}
\end{figure}

The results are summarized in Figs.~\ref{fig_overdamped_small} and \ref{fig_overdamped_big}, corresponding to a small avalanche involving three bond transformations and a larger avalanche comprising six transformations, respectively.
In both figures, panel~(a) shows the sequence of events obtained from numerical continuation, where the hierarchy of bond-level instabilities is ordered from bottom to top.
Panel~(b) presents the corresponding results from the overdamped dynamical simulations.
The vertical axis reports a bond-level activity measure given by the normalized change in bond length between the initial and post-avalanche final configurations.
The color coding is identical in panels~(a) and~(b), facilitating direct comparison between both methods.
The sequence of events observed in the small avalanche shown in Fig.~\ref{fig_overdamped_small}b follows the same hierarchy predicted by numerical continuation in Fig.~\ref{fig_overdamped_small}a.
In contrast, for the larger avalanche (Fig.~\ref{fig_overdamped_big}), the agreement between the two approaches holds only for the first four bond transformations; beyond this point, it breaks down, with the ordering of the final two events reversed.

While both approaches capture the overall structure of bond-level avalanches, the difference observed in the larger avalanche naturally raises the question of which description should be adopted as a reference when resolving the detailed ordering of individual bond transformations.
From our perspective, numerical continuation provides the more faithful representation, as it follows the system along the underlying potential energy landscape and directly identifies the sequence of stability losses associated with successive bond transformations.
By contrast, overdamped dynamics advances the system through explicit time integration, whose outcome may be influenced by finite time stepping, numerical noise, and the dynamical competition between closely spaced instabilities.
Such effects may obscure the fine ordering of individual events, particularly at later stages of an avalanche.

\newpage

\bibliographystyle{apsrev4-2} 
\bibliography{refs.bib} 


\section*{Supplementary material D\\[10pt] Reproducibility and Computational Cost}
$\,$

Here we address two practical aspects of the continuation calculations presented in the main text. First, we discuss the reproducibility of the measured stress--strain curves under small offsets to the initial loading. Second, we discuss the computational cost of the arc-length continuation (AC) method, both as a function of system size and in comparison to standard athermal quasistatic shear/tension (AQS) calculations for the stress--strain curves shown in Fig.~4 of the main text.

\section{Reproducibility under offset initial conditions}
\label{subsec:SI_repro}

\begin{figure*}
    \centering
    \includegraphics[width=0.9\textwidth]{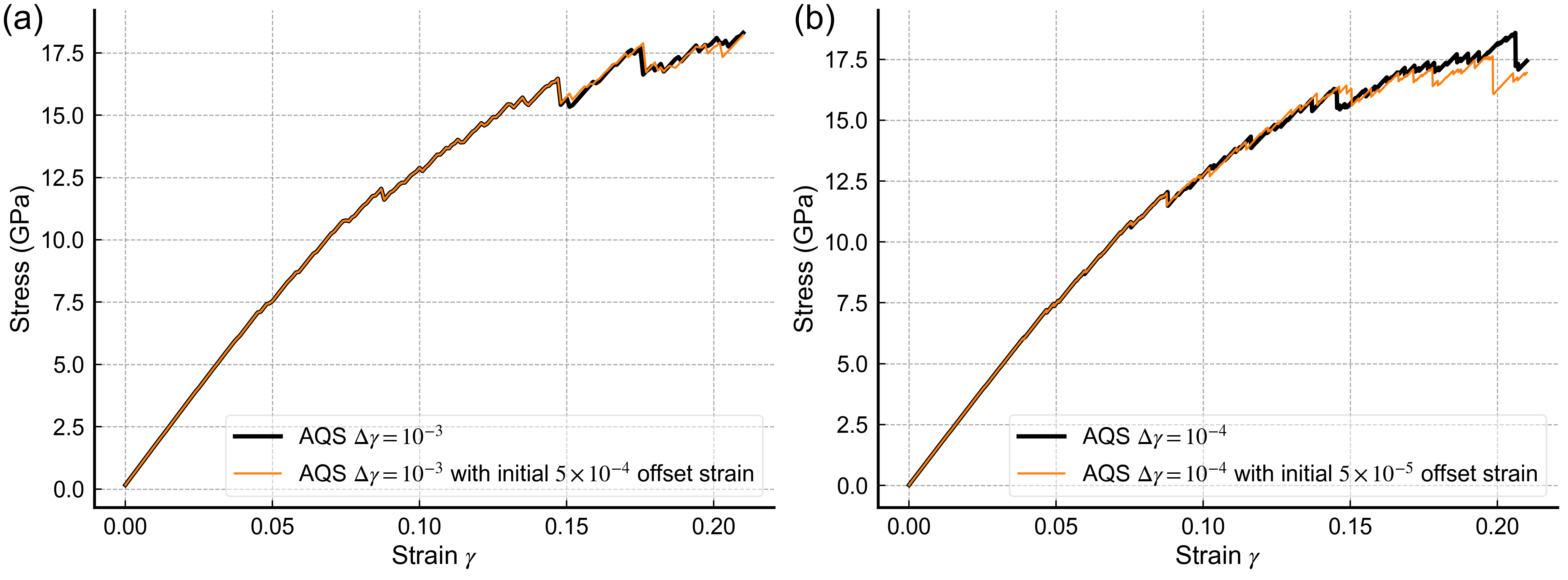}
    \caption{A demonstration of the non-reproducibility of AQS measured stress-strain curves for (a) $\Delta \gamma$ = $10^{-3}$ and (b) $\Delta \gamma$ = $10^{-4}$. In each case, the orange curve represents an AQS simulation with an initial strain offset of $\Delta \gamma/2$. In both cases, the orange and black curve diverge due to the finite step size merging adjacent plastic events.}
    \label{fig:SI_offset_repro}
\end{figure*}

To test reproducibility of the AQS and AC methods, we compared stress--strain curves with those measured using initial configurations that were offset by a small strain increment. For AQS, this means that a calculation performed with strain step $\Delta \gamma$ was repeated from a configuration pre-strained by $\Delta \gamma/2$, and the subsequent loading was then continued using the same step size $\Delta \gamma$. Representative results for $\Delta \gamma = 10^{-3}$ and $\Delta \gamma = 10^{-4}$ are shown in Fig.~\ref{fig:SI_offset_repro} (a) and (b) respectively.

For both AQS step sizes, the stress--strain curves obtained from the offset initial conditions diverge from those obtained from the original initial conditions. This behaviour is expected; in AQS, the imposed strain step defines the resolution with which nearby instabilities are sampled. When plastic events become closely spaced in strain, changing the initial offset by a fraction of the step can alter which instabilities are crossed within a given minimisation, thereby causing physically distinct events to be merged differently. The resulting differences accumulate, leading to distinct stress--strain trajectories and distinct apparent sequences of avalanches.

\begin{figure*}
    \centering
    \includegraphics[width=0.7\textwidth]{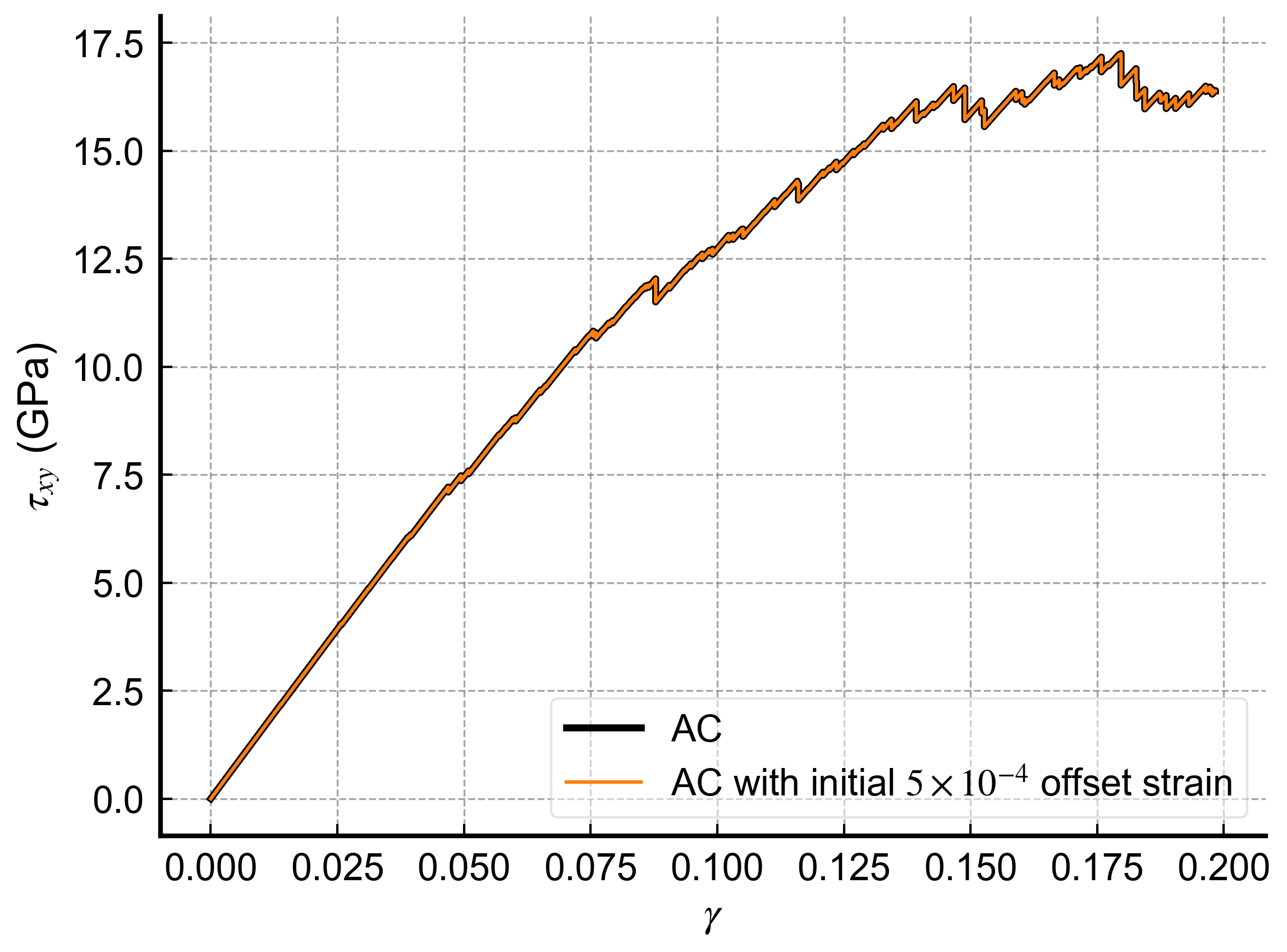}
    \caption{A demonstration of the reproducibility of AC measured stress-strain curves. The orange curve represents a simulation with an initial strain offset of $\delta\gamma = 5\times10^{-4}$. Due to the step-size independence of the method, the orange curve perfectly reproduces the black reference.}
    \label{fig:SI_offset_repro_AC}
\end{figure*}

In principle, because AC follows the equilibrium branch continuously and locates instabilities adaptively, the method should remove the finite-step ambiguity present in AQS simulations. Indeed, we found that provided one uses a strictly dissipative gradient-descent procedure, the AC stress--strain curve obtained from the offset initial configuration is exactly reproduced, including the ordering and decomposition of the plastic events, as shown in Fig.~\ref{fig:SI_offset_repro_AC}. Within the present numerical implementation, this indicates that the avalanche pathways and decompositions identified by AC are unique with respect to small loading offsets, provided that the post-instability relaxation is performed in a reproducible manner.

\section{Scaling and cost}
\label{subsec:SI_scaling_cost}

We now discuss the computational cost of the continuation calculations. The total cost of the method is controlled by three main factors: the cost of evaluating the interatomic potential, the cost of solving the nonlinear continuation equations at each continuation step, and the number of plastic events encountered over the strain interval of interest.

The local ACE machine-learned interatomic potential used in this work has an evaluation cost that scales  linearly with the number of atoms $N$, i.e.\ as $\mathcal{O}(N)$. Since residual evaluations in the continuation calculation are dominated by force evaluations from the potential, this sets the basic cost scaling for both AQS and AC.

\begin{figure*}
    \centering
    \includegraphics[width=0.9\textwidth]{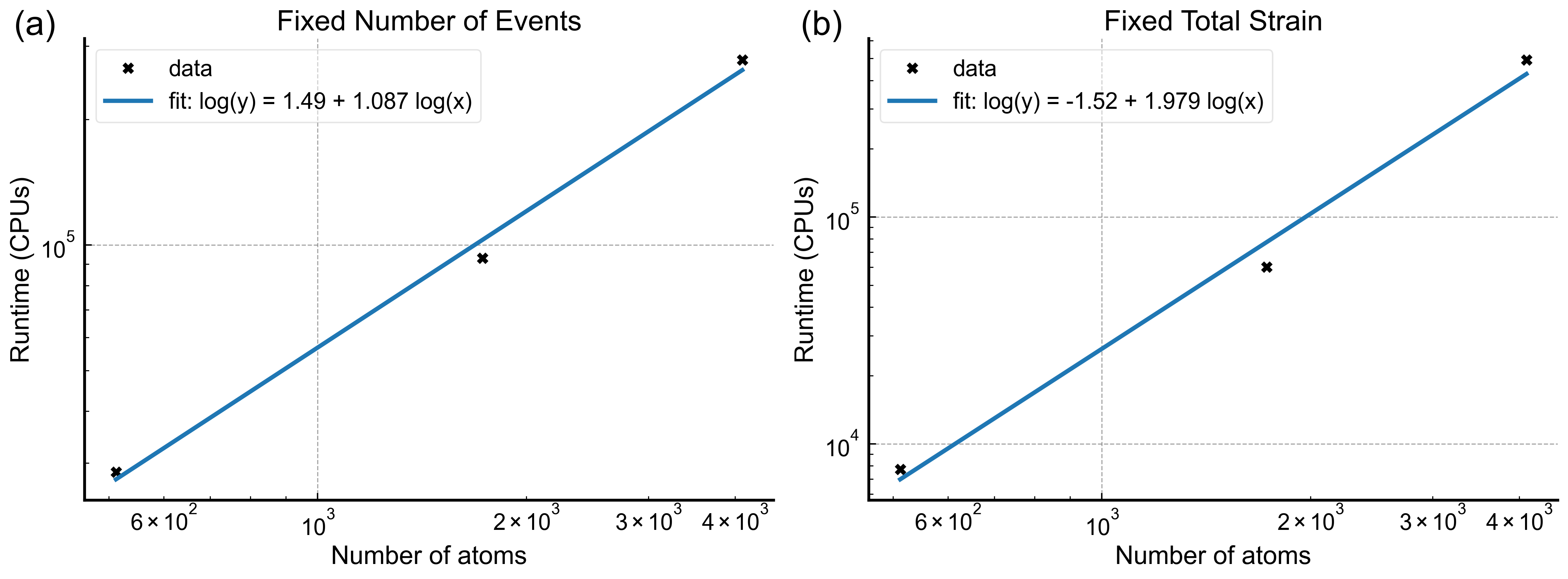}
    \caption{The computational cost scaling of AC simulations with number of atoms $N$ for (a) a fixed number of plastic events (40) and (b) a fixed total strain interval in the plastic flow regime (between 0.1 and 0.2). For a fixed number of plastic events, the scaling was dominated by the cost of evaluating the potential, which was $\mathcal{O}(N)$. For a fixed total strain, as the density of plastic events in the interval also increased with $N$, more AC steps were necessary and the scaling gained an additional factor. We measured the overall scaling to be approximately $\mathcal{O}(N^{2})$.}
    \label{fig:SI_scaling}
\end{figure*}

Each AC step requires the solution of the extended nonlinear system using a Newton--Krylov method. The cost of one such step is given by the product of the number of outer nonlinear iterations and the cost of the inner Krylov iterations. In the present implementation, the number of outer Newton iterations is explicitly capped at six, so the practical scaling is governed primarily by the number of Krylov iterations and residual evaluations required for convergence. If the Krylov iteration count remains approximately independent of system size, then the cost per continuation step is approximately $\mathcal{O}(N)$. Fig.~\ref{fig:SI_scaling} (a) demonstrates this: for a fixed number of resolved plastic events, we measure an overall cost for AC simulations which scales as $\mathcal{O}(N)$.

However, the total cost is not determined only by the cost per step. In the plastic flow regime, the mean strain interval between successive plastic events decreases with increasing system size. Because AC explicitly resolves each event, the total number of continuation steps required to traverse a fixed total strain also increases with $N$. For a fixed total strain, we therefore measure an overall AC cost which is approximately quadratic in system size, $\mathcal{O}(N^2)$; see Fig.~\ref{fig:SI_scaling} (b).

In constrast, in standard AQS if the imposed strain step $\Delta \gamma$ is held fixed as the system size increases, then the nominal total number of AQS minimisations required to reach a fixed total strain is independent of the event density, and the resulting cost scales approximately as $\mathcal{O}(N)$. However, this favourable scaling is obtained only by accepting a fixed strain resolution. As the event spacing decreases with increasing system size, a fixed $\Delta \gamma$ increasingly fails to resolve distinct nearby instabilities, and multiple events are merged within a single strain increment. If instead $\Delta \gamma$ is reduced with system size so as to maintain a comparable event resolution, the total AQS cost correspondingly increases. In this sense, the main practical advantage of AC is not that it has a fundamentally smaller asymptotic scaling exponent, but that it resolves the sequence of instabilities automatically, without requiring an \emph{a priori} choice of strain increment or a manual compromise between cost and event resolution.

As a final test to compare the practical cost of the methods, we measured the total number of ACE potential force evaluations (force calls) required to generate a full stress--strain curve over the domain shown in Fig.~4 in the main text. The number of force calls measured for each simulation are summarised in Table~\ref{tab:SI_cost_compare}. We found that the AC method required $\approx 73 \%$ of the force calls of the $\Delta \gamma = 10^{-4}$ AQS simulation, whilst achieving a significantly better resolution.

\textbf{Hardware details} -- All simulations were run in parallel on 20 AMD 9634 Genoa (Zen 4) 2.25 GHz CPU cores using MPI.

\begin{table}[t]
\centering
\caption{Total number of force calls required to generate the stress--strain curves shown in Fig.~4 of the main text. Note that the AC method requires approximately 73\% of the force calls of the $\Delta \gamma = 10^{-4}$ AQS simulation, and achieves a significantly better resolution.}
\label{tab:SI_cost_compare}
\begin{tabular}{lc}
\hline\hline
Method & Total force calls \\
\hline
AQS, $\Delta \gamma = 10^{-3}$ & $2.0 \times 10^{5}$ \\
AQS, $\Delta \gamma = 10^{-4}$ & $1.3 \times 10^{6}$ \\
AC & $9.5 \times 10^{5}$ \\
\hline\hline
\end{tabular}
\end{table}